 \documentclass[sigconf]{acmart} 

\usepackage{float}
\usepackage{graphicx}
\usepackage{colortbl}
\usepackage{comment}
\makeatletter 
\newcommand*\superimpose[2]{%
  \ooalign{$\m@th#1\@firstoftwo#2$\cr
           \hidewidth$\m@th#1\@secondoftwo#2$\hidewidth}%
}
\makeatother 
\newcommand*\threedotsord{\mathpalette\superimpose{{\mathop:}{\cdot}}} 
\newcommand*\threedotsopen{\mathopen{\threedotsord}}   

\AtBeginDocument{%
  \providecommand\BibTeX{{%
    \normalfont B\kern-0.5em{\scshape i\kern-0.25em b}\kern-0.8em\TeX}}}

\copyrightyear{2021}
\acmYear{2021}
\setcopyright{acmlicensed}\acmConference[CHI '21]{CHI Conference on Human Factors in Computing Systems}{May 8--13, 2021}{Yokohama, Japan}
\acmBooktitle{CHI Conference on Human Factors in Computing Systems (CHI '21), May 8--13, 2021, Yokohama, Japan}
\acmPrice{15.00}
\acmDOI{10.1145/3411764.3445467}
\acmISBN{978-1-4503-8096-6/21/05}

\begin{document}

\title{How the Design of YouTube Influences User Sense of Agency}



\author{Kai Lukoff}
\affiliation{%
  \institution{University of Washington}
}
\email{kai1@uw.edu}

\author{Ulrik lyngs}
\affiliation{%
  \institution{University of Oxford}
}
\email{ulrik.lyngs@cs.ox.ac.uk}

\author{Himanshu Zade}
\affiliation{%
  \institution{University of Washington}
}
\email{himanz@uw.edu}

\author{J. Vera Liao}
\affiliation{%
  \institution{University of Washington}
}
\email{ljzj@uw.edu}

\author{James Choi}
\affiliation{%
  \institution{University of Washington}
}
\email{jchoi408@uw.edu}

\author{Kaiyue Fan}
\affiliation{%
  \institution{University of Washington}
}
\email{kaiyuef@uw.edu}

\author{Sean A. Munson}
\affiliation{%
  \institution{University of Washington}
}
\email{smunson@uw.edu}

\author{Alexis Hiniker}
\affiliation{%
  \institution{University of Washington}
}
\email{alexisr@uw.edu}

\renewcommand{\shortauthors}{Lukoff, et al.}



\begin{abstract}
  In the attention economy, video apps employ design mechanisms like autoplay that exploit psychological vulnerabilities to maximize watch time. Consequently, many people feel a lack of agency over their app use, which is linked to negative life effects such as loss of sleep. Prior design research has innovated \textit{external mechanisms} that police multiple apps, such as lockout timers. In this work, we shift the focus to how the \textit{internal mechanisms} of an app can support user agency, taking the popular YouTube mobile app as a test case. From a survey of 120 U.S. users, we find that autoplay and recommendations primarily undermine sense of agency, while search and playlists support it. From 13 co-design sessions, we find that when users have a specific intention for how they want to use YouTube they prefer interfaces that support greater agency. We discuss implications for how designers can help users reclaim a sense of agency over their media use.
\end{abstract}



\maketitle

\section{Introduction}
\textit{$``$At Netflix, we are competing for our customers’ time, so our competitors include Snapchat, YouTube, sleep, etc.$"$ } \\ - Reed Hastings, Netflix CEO \cite[p.50]{Williams2018-co} 
\vspace{\baselineskip}

\noindent In the attention economy, social media apps employ a variety of design mechanisms--such as eye-catching notification icons, tempting clickbait, and never-ending autoplay--to maximize their share of the user's time. In this pursuit, designers and tech industry insiders warn that many of these mechanisms exploit psychological vulnerabilities and harm the interests of the user \cite{Lewis2017-dr,Burr2018-ut}.


It is no accident then that social media use is often associated with a loss of sense of agency \cite{Baumer2018-gv}. People self-report that their desire to consume media frequently conflicts with their plans or goals and that they fail to resist about three-quarters of the time \cite{Delaney2017-ld}. And loss of control is a key component of many measures of problematic technology use \cite{Cheng2019-me}. \par

In response, digital wellbeing researchers have innovated what we term \textit{external mechanisms} that help users manage or monitor their app use, such as lockout timers \cite{Kim2019-pq} and productivity dashboards \cite{Kim2016-pj}. While these mechanisms apply universally to many different apps, they do not change the \textit{internal mechanisms} within an app, such as autoplay, that might lead it to be problematic in the first place.

One promising approach is to redesign these mechanisms for a greater sense of agency, i.e., an individual’s experience of being the initiator of their actions in the world \cite{Synofzik2008-ia}. Low sense of agency over technology use is associated with negative life impacts such as a loss of social opportunities, productivity, and sleep \cite{Caplan2010-ox} that often motivate digital wellbeing efforts to begin with. Moreover, a lack of sense of agency itself can be understood as a driver of the dissatisfaction that people often feel with their social media use \cite{Marino2018-qb}. 


In this work, we take the mobile app for YouTube, the most widely used social media service in the United States \cite{Perrin2019-jr}, as a test case to understand and redesign how internal mechanisms influence sense of agency. The design of YouTube must balance the interests of many different stakeholders. For example, policymakers may wish to exert control over extremist content. Advertisers may wish to control how much time users spend on ads. Designers may wish to control how much time users spend in the app. Content creators may wish to control how much time users spend on their channel. All of these stakeholders merit consideration, however, in this work we focus specifically on \textit{users} and how design influences the control they feel over the time they spend in the mobile app.

We investigate two research questions in two studies that build upon each other:\par

\begin{itemize}
	\item \textbf{RQ1: What existing mechanisms in the YouTube mobile app influence sense of agency?}\par

In a survey, we asked 120 YouTube users which mechanisms make them feel most and least in control of how they spend their time in the YouTube mobile app. \par

	\item \textbf{RQ2: What changes to these mechanisms might increase sense of agency?}

Based on the responses to the survey, we redesigned four internal mechanisms to change user sense of agency in the YouTube app: recommendations, playlists, search, and autoplay. In co-design sessions, we then asked 13 YouTube users to sketch changes of their own and evaluate our mockups. We also asked how much control they would prefer to have in different situations.\par

\end{itemize}\par

\noindent The two contributions of this work are:\par

\begin{enumerate}
	\item We identify the internal design mechanisms that influence users' sense of agency over how they spend time in the YouTube mobile app and how they might be changed. While some of these mechanisms are expected (e.g., autoplay), others are less so (e.g., playlists) and suggest promising directions for digital wellbeing (e.g., designing to support ‘microplans’ that guide behavior within a single session of use).\par

	\item We distinguish when designing for a sense of agency is desirable from when it might actually go against what users want. Participants in our co-design sessions preferred greater control when they had a specific intention for using the app (e.g., to cook a recipe) than when they had a non-specific intention (e.g., to relax), in which case they wanted to let the app take control. We propose ways for designers to navigate this mixed preference for different levels of control at different times.
\end{enumerate}\par

\section{Background and Motivation}

\subsection{Designing to Undermine Sense of Agency}
Design practitioners have raised concerns about dark patterns, interfaces that are designed to manipulate a user into behavior that goes against their best interests \cite{Gray2018-zz,Lukoff2021-ru}. Brignull's original types of dark patterns focus on financial and privacy harms to the user \cite{Brignull_undated-pv}. However, given that people routinely report using technology in ways that are a waste of their time and that they later regret \cite{Ko2015-yc,Hiniker2016-cs,Lukoff2018-km,Ames2013-ja}, there is a need for research to examine which design patterns prompt such \textit{attentional harms} for the user. We might term these attention capture dark patterns, designs that manipulate the user into spending time and attention in an app against their best interests. \par

Tech industry insiders, like the ex-President of Facebook, warn that social media apps are especially likely to innovate and employ design patterns that \textit{"consume as much of your time and conscious attention as possible"} \cite{Pandey2017-it}. For social games, one such a proposed pattern is \textit{$``$playing by appointment,$"$} wherein a player must return to play on a schedule defined by the game, or else lose their precious resources \cite{Zagal2013-nm}. For social media, a common suggestion in popular self-help guides is to take back control by turning off notifications \cite{noauthor_undated-zy,Kamenetz2018-sl}. However, it is not yet established that these mechanisms are the ones that lead users to feel a loss of control. For example, some users report that notifications actually reduce their checking habits, since they know they will be alerted when their desired content is ready \cite{Oulasvirta2012-cc}. \par

YouTube is an important case for better understanding the design mechanisms of attention capture. YouTube has over two billion monthly users worldwide \cite{YouTube_undated-js} and is extremely popular in the U.S., where about three-quarters of adults report using YouTube on their smartphone, with 32$\%$  using it several times a day, 19$\%$ about once per day, and 49$\%$ less often \cite{Perrin2019-jr}. It is also frequently reported as a source of distraction \cite{Aagaard2015-rf}, suggesting that it is a good site for the investigation of attention capture dark patterns. In particular, Youtube's algorithmic recommendations merit special consideration as they drive more than 70$\%$ of watchtime \cite{Solsman2018-wn}. \par

\subsection{Designing to Support Sense of Agency}
Reducing screentime in certain apps is a common measure of success in digital wellbeing tools. The two most popular mobile operating systems, Android and iOS, both come pre-installed with tools for the user to track and limit their time in mobile apps. Within the YouTube app itself, there are also features to manage time spent: `Time watched statistics,' which shows how much time a user has spent on YouTube in each of the last 7 days, and the `Take a break reminder,' which periodically prompts the user to take a rest. A strength of addressing digital wellbeing via such screentime tools is that time spent is easy to track and easy to understand.

However, a weakness of this approach is that reducing screentime is often a poor proxy for what users actually want. Instead, user intentions are often highly specific, such as wanting to reduce the time spent on targeted features of an app (e.g., on the Facebook newsfeed, but not in Facebook groups) or in certain contexts (e.g., when with family, but not when commuting on the bus) \cite{Lukoff2018-km,Lyngs2020-pm,Hiniker2016-cs}. 

Within YouTube, there are two digital wellbeing features that do move beyond time spent controls and offer more granular control. The `Notifications digest' lets a user bundle push notifications together into a single notification each day, which may reduce the triggers that lead to non-conscious, habitual use \cite{Lyngs2018-qi}. `Autoplay toggle' lets a user decide to stop the next video from playing automatically; this may preserve the natural stopping point that comes at the end of the video, a mechanism that has been shown to help users set more deliberate boundaries around use \cite{Hiniker2018-ra}. While the notification digest and the autoplay toggle clearly do more than just track and limit time, it is not immediately clear by what measure of success they might be evaluated.

One promising alternative to the screentime paradigm is to design for \textit{sense of agency}, the focus of this paper. Sense of agency is a construct that refers to an individual’s experience of being the initiator of their actions in the world \cite{Synofzik2008-ia}. Sense of agency can be broken down into \textit{feelings of agency}, that is, the in-the-moment perception of control, and \textit{judgments of agency}, that is, the post hoc, explicit attribution of an action to the self or other \cite{Synofzik2008-ia}. In the present paper, we focus on the latter, judgments of agency.

Sense of agency matters for digital wellbeing in at least three ways. First, supporting user control is a common principle in HCI design guidelines \cite{Coyle2012-qr,Nielsen1994-ss,Shneiderman2004-yh}. Designing for an \textit{$``$internal locus of control$"$} is one of Shneiderman and Plaisant’s  Eight Golden Rules of Interface Design, arising from the observation that users want \textit{$``$the sense that they are in charge of an interface and that the interface responds to their actions$"$} \cite{Shneiderman2004-yh}. Second, a low sense of control over technology use predicts greater negative life effects, e.g., internet use leading to missed social activities \cite{Caplan2010-ox} and smartphone use leading to the loss of a career opportunity or significant relationship \cite{Jeong2016-cg}. Scales of problematic technology use generally measure both (a) lack of control and (b) negative life impacts, suggesting that ‘the problem’ is a combination of these two factors \cite{Cheng2019-me,Cash2012-wj}. Third, and perhaps most importantly, sense of agency matters in its own right. Feeling in control of one’s actions is integral to autonomy, one of the three basic human needs outlined in self-determination theory \cite{Ryan2006-lb}. More specific to technology use, it is also central to user (dis)satisfaction with smartphones \cite{Davis2019-au,Harmon2013-ja} and Facebook use \cite{Cheng2019-me,Marino2018-qb}. \par

Prior work has investigated different ways that interfaces can support sense of agency. First, some input modalities seem to support a greater sense of agency than others (e.g., keyboard input versus voice commands) \cite{Limerick2015-ui}. Second, a system's feedback should match a user's predicted feedback \cite{Limerick2014-lk}. Third, a study of flight navigation systems found that increasing the level of automation reduced sense of agency \cite{Berberian2012-gt}. These lessons might be revisited in the domain of digital wellbeing, as how an interface modulates sense of agency may vary with context \cite{Limerick2014-lk}.



\subsection{Design Mechanisms for Digital Wellbeing}

\begin{figure*}
    \centering
    \includegraphics[scale=0.7]{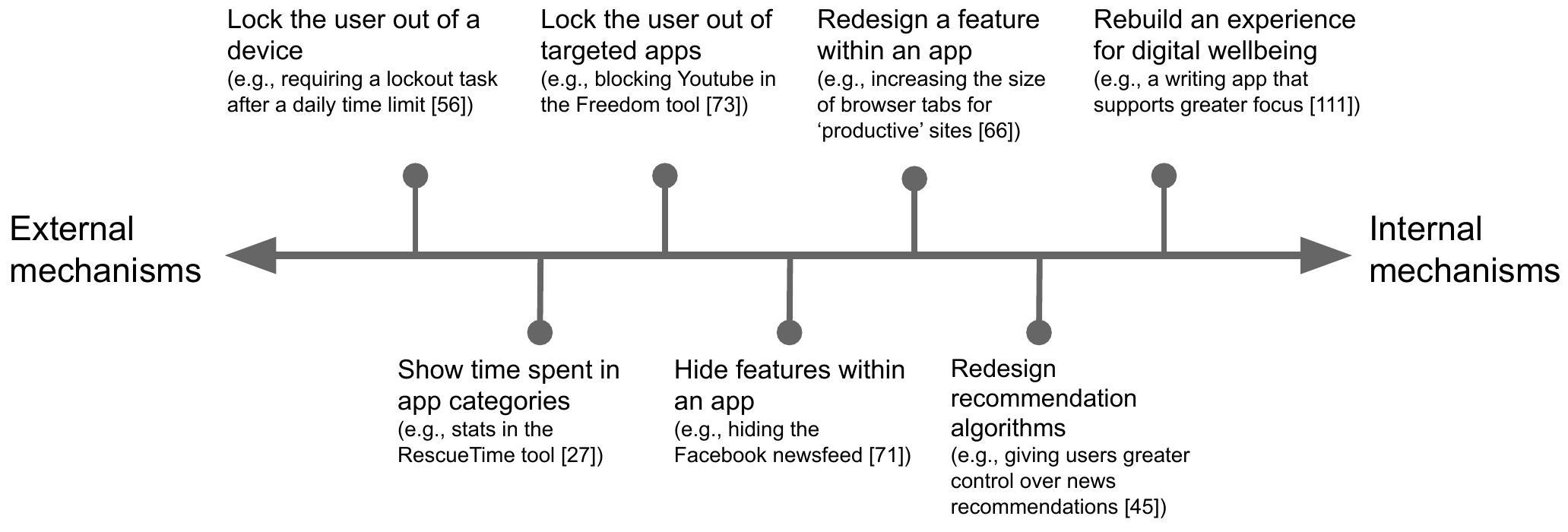}
    \caption{Mechanisms that influence how people spend their time in apps can be placed along a spectrum, as in these examples. External mechanisms monitor or police apps, while internal mechanisms redesign or rebuild the experience within a problematic app. Internal mechanisms offer designers a more targeted way of supporting user agency.}
    \label{fig:fig1}
\end{figure*}

The mechanisms\footnote{We use the term $``$mechanism$"$  to describe one component of a larger design (although some digital wellbeing designs do consist of a single mechanism)} of digital wellbeing interventions can be placed along a spectrum (see \textbf{Figure \ref{fig:fig1}}). At one end are external mechanisms that monitor or police apps, such as screentime statistics and lockout timers. A hallmark of an external mechanism is that it functions identically across multiple apps, as in a timer that locks the user out of social media, gaming, \textit{and} video apps. However, external mechanisms do not significantly change the experience \textit{within} individual apps.\par

At the other end of the spectrum, internal mechanisms contribute to the redesign or rebuild of an experience. For example, Focus Mode in \textit{Microsoft Word} redesigns the writing process by hiding all formatting options \cite{Baab-Muguira2017-jr}. Going a step further, the standalone app \textit{Flowstate}\ not\ only\ offers a minimal interface, but also deletes all text on the page if the user stops writing for longer than seven seconds \cite{Statt2016-uo}. Internal mechanisms fundamentally change the experience within a problematic app, or rebuild it into a new experience entirely. \par

At present, design researchers have innovated many tools on the external side of the spectrum, that monitor and police multiple apps in the same way \cite{Kim2019-pq,Kim2019-cd,Collins2014-bv,Monge_Roffarello2019-xt,Okeke2018-ad}. Likewise, industry designers have built tools that apply the same time lockout mechanism to all apps, such as the screentime tools that come pre-installed on Android and iOS.

In contrast to external mechanisms, the space of internal mechanisms is relatively underexplored (see \cite{Lottridge2012-tj,Harambam2019-qu} for notable exceptions), but holds particular promise for increasing user agency in two ways. First, designers can craft more targeted interventions with internal mechanisms than with external ones. External mechanisms, such as locking the user out of a device, often require sacrifices that users are reluctant to accept \cite{Tran2019-lv,Kim2019-pq}. Whereas an external mechanism might block the Facebook app after time is up, a more internal could reconfigure the newsfeed to show only content from close personal friends. A redesign of internal mechanisms may be able to remove problematic aspects from an app, while still retaining its benefits.\par

Second, internal mechanisms shift the focus from fighting distractions to aligning interests. External mechanisms often respond to the temptations of problematic apps with microboundaries \cite{Cox2016-ts} or restraints on interactions \cite{Park2018-xs}. However, this sets up an arms race in which the designers of digital wellbeing tools are always in a defensive position. An alternative is for designers to reenvision the internal mechanisms that lead to compulsive use in the first place \cite{Tran2019-lv}. Looking at the mechanisms inside of specific apps may encourage designers to not just block existing mechanisms but to innovate better ones, such as \textit{Flowstate’s} seven seconds rule for writing. This paper presents an examination how such internal mechanisms can be redesigned to support sense of agency.

\section{Study 1: Survey of 120 YouTube users}
Study 1 examines how existing mechanisms in the YouTube mobile app support or undermine sense of agency (\textbf{RQ1}). We decided to start by investigating user's experiences in the current app before proceeding to design and evaluate potential changes in Study 2 (\textbf{RQ2}). Both studies were approved by the University of Washington's Institutional Review Board.

\subsection{Participants}
\subsubsection{Recruitment.} To obtain a general sample of users of the YouTube mobile app, we recruited from Amazon Mechanical Turk workers in the United States. Participants were invited to \textit{$``$Help us     understand how people spend their time on the YouTube mobile app.$"$ } They were required to meet four inclusion criteria: \par
\begin{enumerate}
	\item A task approval rating greater than 98$\%$  for their prior work on Mechanical Turk, indicating a history of high-quality responses.\par

	\item Own a smartphone. Three members of our research team tested the YouTube mobile app on both Android and iPhone and found that the app has nearly identical features and only minor stylistic differences, so we accepted users of both types of devices as participants (80 Android, 40 iPhone users).\par

	\item Spend a minimum of 3 hours on YouTube in the past week (across all devices), according to their time watched statistics in the YouTube app. In the survey, participants saw instructions with screenshots that showed where to find this statistic in the app, confirmed that they had found it, and then entered it into the survey. To see time watched statistics, users must be signed into the app. \par

	\item Of the time they spend on YouTube, 20$\%$  or more is on their smartphone (self-estimated). 
\end{enumerate}\par

\subsubsection{Demographics.} A total of \textbf{120 participants} met the inclusion criteria and completed the survey (see demographics in \textbf{Table \ref{tab:tab1}}). We excluded responses from an additional 7 participants who started but did not complete the survey. We oversampled men, Asians, and young people relative to the 2019 estimates of the United States Census Bureau \cite{United_States_Census_Bureau_undated-re}. Other participant samples may use the YouTube mobile app differently, e.g., users in emerging countries for whom a smartphone is often their only device for watching videos \cite{Silver2019-xj}. Further research is required to determine whether our results apply to other populations. \par

\begin{table}
\centering
\huge 
\resizebox{\linewidth}{!}
{%
\begin{tabular}{p{0.2\textwidth} p{0.65\textwidth}}
\hline
Gender identity           & Man (63\%), Woman (36\%), Non-binary (0\%), Prefer not to say (1\%)                                              \\ \hline
Age range                 & 18-24 (8\%), 25-34 (41\%), 35-44 (40\%), 45-54 (11\%), 55+ (1\%)                                 \\ \hline
Education                 & High school (22\%), Associate degree (22\%), Bachelor’s degree (46\%), Advanced degree (11\%)   \\ \hline
Household income (US)    & <25K (14\%), 25-50K (23\%), 50-75K (30\%), 75-125K (20\%), > 125K (11\%), prefer not to say (2\%) \\ \hline
Race (choose one or more) & White (69\%), Asian (17\%), Black (9\%), Hispanic/Latino (4\%), Native American (2\%)             \\ \hline
\end{tabular}%
}
\caption{Demographics of the 120 survey participants}
\label{tab:tab1}
\end{table}

\subsubsection{YouTube use.} 
Participants spent a median of 101 minutes per day (interquartile range: 57-156) on YouTube across all devices in the week prior to the survey. Of this time, participants estimated they spent a median of 50$\%$ (interquartile range: 30-75$\%$) in the mobile app. For comparison, the YouTube press page states that mobile accounts for over 70$\%$ of watchtime \cite{YouTube_undated-js}. Upon multiplying these two responses together for each participant, we found that participants spent an average of 70 minutes per day in the YouTube mobile app. This is similar to the average for all YouTube users: in 2017, YouTube shared that signed-in users spend an average of more than 60 minutes per day in the mobile app \cite{Matney2017-ac}. We neglected to ask whether participants were using the paid YouTube premium service, which removes ads and can play videos offline and in the background; however, Google reports that only 1$\%$  of YouTube’s monthly visitors subscribe to this service \cite{Spangler_undated-my}.\par

\subsection{Procedure}
Participants answered questions in an online survey. The initial questions asked about our four inclusion criteria. Eligible participants continued on to background questions about their demographics and YouTube use. The complete survey wording, along with all of the other appendices for this study can be found at: \href{https://osf.io/w3hmd}{https://osf.io/w3hmd} \par

To investigate \textbf{RQ1}, one question table asked about things that made participants feel \textit{most in control} of how they spend their time on YouTube (See \textbf{Table \ref{tab:tab2}}). A second question table asked about things that made them feel \textit{less in control}. The order of these two question tables was randomized. In terms of wording, we chose to ask about feeling \textit{"in control,"} as this is how sense of agency has been measured in previous studies of sense of agency in HCI (e.g., \cite{Metcalfe2007-cp}) and on a self-report scale \cite{Tapal2017-lk}. We used the informal term \textit{$``$things$"$} because, in piloting the survey, we found that testers were unsure about whether certain things (e.g., recommendations and ads) counted as \textit{$``$mechanisms$''$} of the app and we did not want to provide examples that would bias responses. In total, each participant was required to submit 6 responses for things that influenced their sense of agency on YouTube (3 for most in control, 3 for least in control). \par

\begin{table*}
\small
\centering

\begin{tabular}{|l|l|l|}
\hline
        & \begin{tabular}[c]{@{}l@{}} \underline{Thing Question:} What are 3 things about\\ the mobile app that lead you to feel most\\ in control over how you spend your time \\ on YouTube?\end{tabular} & \begin{tabular}[c]{@{}l@{}}\underline{Explain Question:} How does this thing\\ make you feel more in control of how you\\ spend your time on YouTube?\end{tabular}                                                                                        \\ \hline
Thing 1 & \textit{\begin{tabular}[c]{@{}l@{}}“I am able to quickly access my subscribed\\  channels.”\end{tabular}}                                                                           & \textit{\begin{tabular}[c]{@{}l@{}}“I don’t spend uncontrolled amounts of time\\ browsing through videos that may or may not\\ be related to what I want to watch.”\end{tabular}}                                                             \\ \hline
Thing 2 & \textit{\begin{tabular}[c]{@{}l@{}}“I am able to get notifications of certain\\ channels or videos getting posted.”\end{tabular}}                                                   & \textit{\begin{tabular}[c]{@{}l@{}}“I will know exactly when a new video goes\\ up that I may be interested in watching.\\ This way I am not randomly checking for\\ uploads and spending extra time searching\\ and browsing.”\end{tabular}} \\ \hline
Thing 3 & \textit{“Screen/watch time.”}                                                                                                                                                       & \textit{\begin{tabular}[c]{@{}l@{}}“I can follow trends and tell when I am\\ spending more time than usual on the app.”\end{tabular}}                                                                                                         \\ \hline
\end{tabular}%

\caption{The wording and format of the \textit{“more in control”} question in the survey. The example responses here come from a single study participant. All participants also completed a second version of this question table, with the text modified from \textit{“most”} to \textit{“least”} in the \underline{Thing Question} and from \textit{“more”} to \textit{“less”} in the \underline{Explain Question}.}
\label{tab:tab2}
\end{table*}

Participants were compensated $\$$6.00 for answering all questions, an amount that exceeds the U.S. minimum wage ($\$$7.25 per hour). The survey took a median of 21 minutes to complete (interquartile range: 15-29). \par

\subsection{Coding reliability thematic analysis}
We conducted a coding reliability thematic analysis \cite{Boyatzis1998-wf,Braun2018-th}, in which we first established reliable codes for design mechanisms and then used them to generate themes that captured shared meanings. We started by iteratively coding the 720 responses (6 per participant). Each \textit{$``$thing$"$}  was analyzed as a single response, combining answers to the Thing Question and the Explain Question (i.e., one row in \textbf{Table \ref{tab:tab2}}). In our first pass, two researchers individually reviewed all responses and met to develop initial codes. At this stage, we eliminated 112 responses without any substantive content, e.g., \textit{``I can’t think of anything else.''} Of the 112 responses without substance, 55 came from ``less in control'' and 57 from ``more.''

We further limited coding to responses that specified a mechanism within the interface of the YouTube mobile app, i.e., something the app’s designers could directly change. This included responses such as, $``$\textit{Recommended videos} - \textit{Being shown recommended videos is like a moth to a light for me,$"$} which was coded as ‘recommendations’. It excluded responses about situational factors that are largely outside of the control of the designer such as, \textit{$``$I make my own decisions - I am a conscious person who can make decisions on what I do.$"$} This eliminated 141 more responses (59 from ``less in control” and 82 from ``more in control''). Interestingly, ``more in control'' included 28 responses that we coded as willpower, e.g., \textit{$``$I make my own decisions,$"$} with only 1 such response for ``less”. This suggests a potential self-serving bias \cite{forsyth2008self} wherein in-control behavior is attributed to one's own willpower whereas out-of-control behavior is attributed to external factors. The other responses that we removed were about characteristics of mobile phones (e.g., \textit{``The app is easy to access and tempt me on my phone...”}) and usability issues (e.g., \textit{``it crashes on me every other day or so''} and \textit{``it consumes a lot of battery life''}) that are not specific to the interface of the YouTube mobile app. After excluding these responses, we continued with coding the 467 responses that referenced a specific design mechanism.\par

In our second pass, we applied the initial codes to 120 randomly selected responses and met to discuss. Since one mechanism (recommendations) came up more often than all others, we developed three subcodes for how recommendations affected participant experiences on YouTube. After merging similar codes, our codebook consisted of 21 design mechanisms, such as autoplay, playlists, and multiple device sync. In our third pass, we each independently coded the same 50 randomly selected responses. Interrater reliability was assessed using Cohen’s kappa, with $ \kappa $  = 0.73 indicating substantial agreement \cite{Landis1977-xl}. In our fourth pass, we each coded half of the remaining responses, discussed the final counts, and selected several representative quotes for each code. The first author then wrote up a draft of the coding results and reviewed together with the other authors. We mapped codes (design mechanisms) to potential themes, generating three higher-level themes that structured our final writeup. In our analysis and writeup, we noted cases where responses for an individual code were split with regards to a theme, e.g., ‘notifications’ sometimes supported and sometimes undermined ‘planning ahead’.

\subsection{Results and Analysis}
\subsubsection{Design Mechanisms.} 
\begin{figure}
    \centering
    \includegraphics[scale=0.15]{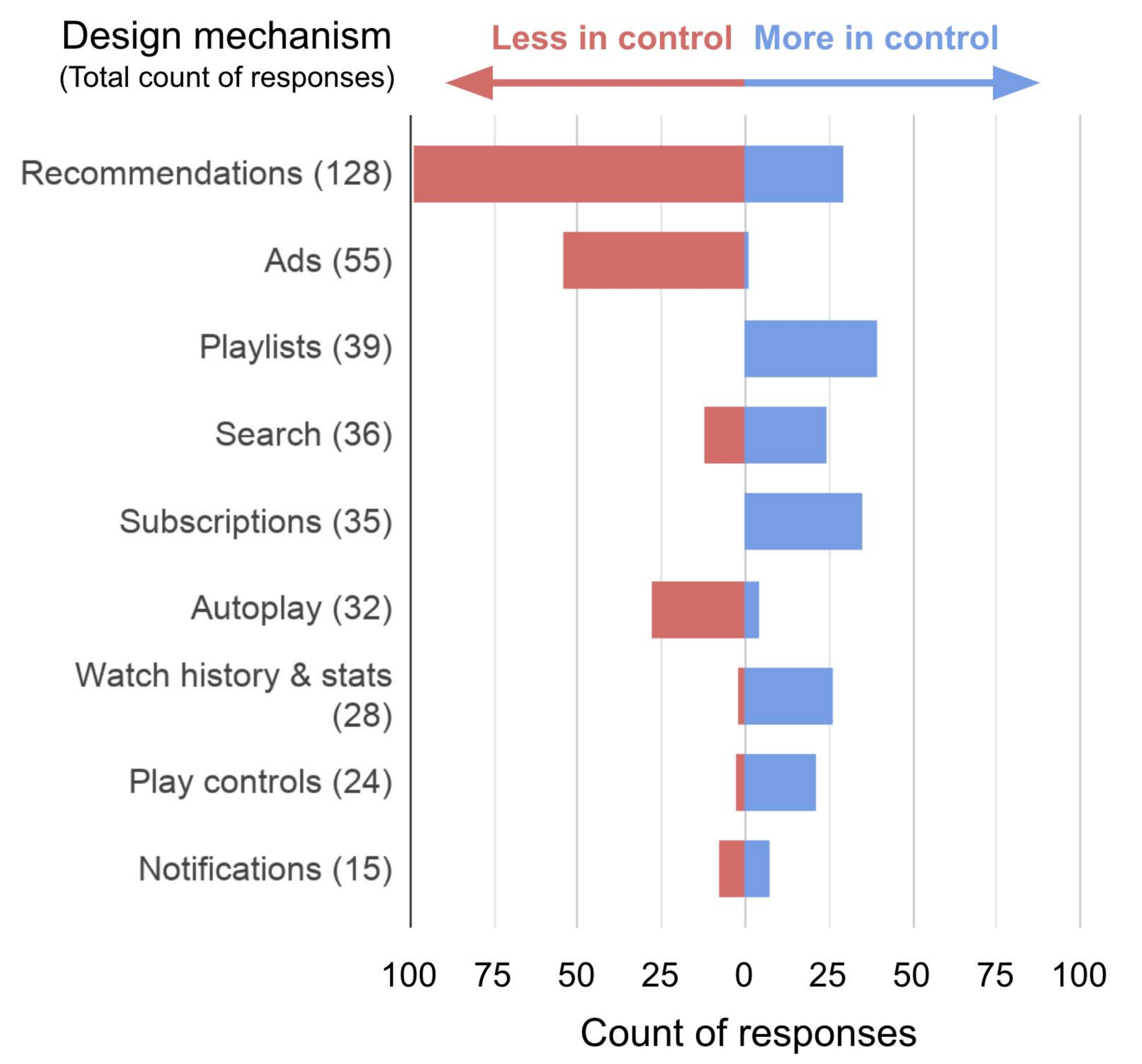}
    \caption{This diverging bar chart shows how many times these nine design mechanisms led participants to feel more control or less control. Recommendations, ads, and autoplay primarily made respondents feel less in control. Playlists, search, subscriptions, play controls, and watch history $\&$  stats primarily made respondents feel more in control. Notifications were sometimes mentioned as leading to more control and sometimes to less.}
    \label{fig:fig2}
\end{figure}

467 responses referenced a specific design mechanism (246 for less in control, 221 for more in control). Nine mechanisms were described as influencing sense of agency 15 or more times and are the focus of our analysis.\footnote{ Mechanisms mentioned 15 or more times covered 392 of 467 responses (84$\%$) that referenced a design mechanism. Mechanisms mentioned fewer than 15 times included content moderation (12), playing videos in the background (12 responses), syncing across multiple devices (9), comments (9), ratings (8), and YouTube’s ‘Take a break reminders' (5). The 6 remaining mechanisms were mentioned fewer than 5 times each.} \textbf{Figure \ref{fig:fig2}} provides a glanceable view of how many times each of these nine mechanisms was mentioned as leading to more or less control. \textbf{Table \ref{tab:tab3}} shows the same data with a description and example response for each mechanism. \textbf{Appendix I} contains annotated screenshots that show the exact implementation of these nine mechanisms in the YouTube mobile app as they appeared when participants provided their feedback. \par

In summary, recommendations were the most frequently mentioned mechanism, accounting for 27$\%$  of all responses. Recommendations, ads, and autoplay primarily made respondents feel less in control. Playlists, search, subscriptions, play controls, and watch history $\&$  stats primarily made respondents feel more in control. Notifications were divided with about half of responses in each direction.\par 
\vspace{\baselineskip}

\begin{table*}
\centering
\resizebox{\textwidth}{!}{%
\begin{tabular}{|c|c|c|c|c|}
\hline
\rowcolor[HTML]{A4C2F4}
\textbf{\begin{tabular}[c]{@{}c@{}}Design\\ Mechanism\end{tabular}}                    & \textbf{Description}                                                                                                                                                                        & \textbf{\begin{tabular}[c]{@{}c@{}}Count\\ of re-\\ sponses\end{tabular}} & \begin{tabular}[c]{@{}c@{}}\textbf{Less in} \\ \textbf{control} (\%\\ of responses)\end{tabular} & \begin{tabular}[c]{@{}c@{}}\textbf{Representative quote(s)}\\ (2 quotes if minority opinion on\\ direction of control >20\% of responses)\end{tabular}                                                                                                                                                                                           \\ \hline
\begin{tabular}[c]{@{}c@{}}Recommendations...\\ (see 3 subcodes \\ below)\end{tabular} & \begin{tabular}[c]{@{}c@{}}Recommended videos on the home, \\ explore, \& video player screens.\end{tabular}                                                                                    & 128                                                                       & 77\%                                                                                  & See subcodes in the 3 rows below.                                                                                                                                                                                                                                                                                                    \\ \hline
\begin{tabular}[c]{@{}c@{}}/ Irrelevant \\recommendations\end{tabular}                & \begin{tabular}[c]{@{}c@{}}Repetitive, dull, or\\ generic recommendations that the\\ user is not interested in.\end{tabular}                                                         & \begin{tabular}[c]{@{}c@{}}42 \\ (of 128)\end{tabular}                     & 100\%                                                                                 & \textit{\begin{tabular}[c]{@{}c@{}}“The related videos are sometimes videos I’ve seen before,\\ over and over.”\end{tabular}}                                                                                                                                                                                                        \\ \hline
\begin{tabular}[c]{@{}c@{}}/ Relevant \\recommendations\end{tabular}                 & \begin{tabular}[c]{@{}c@{}}Engaging or catchy recommenda-\\ tions that the user is interested in.\end{tabular}                                                                              & \begin{tabular}[c]{@{}c@{}}45 \\(of 128)\end{tabular}                      & 53\%                                                                                  & \textit{\begin{tabular}[c]{@{}c@{}}“YouTube has very good algorithms that know what I like,\\ when I want it.” —VS.— “I have a hard time not looking\\at the suggested videos that the algorithm picks for me...\\ I almost always justify watching just one more video.”\end{tabular}}                        \\ \hline
\begin{tabular}[c]{@{}c@{}}/ Customization \\ settings\end{tabular}                                                                    & \begin{tabular}[c]{@{}c@{}}Settings to customize location, \\quantity, or content of \\recommendations.\end{tabular}                                                                      & \begin{tabular}[c]{@{}c@{}}41 \\(of 128)\end{tabular}                      & 81\%                                                                                  & \textit{\begin{tabular}[c]{@{}c@{}}“Not having control over the trending list. I feel like I’m\\ force-fed content.”\end{tabular}}                                                                                                                                                                                                   \\ \hline
\rowcolor[HTML]{D9D9D9}
Ads                                                                                    & \begin{tabular}[c]{@{}c@{}}Ads that appear before, during, and\\ after videos in the player.\end{tabular}                                                                                   & 55                                                                        & 98\%                                                                                  & \textit{\begin{tabular}[c]{@{}c@{}}“I feel as if I am forced to watch ads, which can suck up\\ a lot of time.”\end{tabular}}                                                                                                                                                                                                         \\ \hline
\begin{tabular}[c]{@{}c@{}}Playlists (includes\\ Watch Later)\end{tabular}              & \begin{tabular}[c]{@{}c@{}}Creating, saving, and playing a list\\ of videos. Watch Later is a default\\ playlist for all users. Playlists \\autoplay all videos on the list.\end{tabular} & 39                                                                        & 0\%                                                                                   & \textit{\begin{tabular}[c]{@{}c@{}}“I can create playlists or queue videos in advance to limit\\ what I watch to a specific list instead of endlessly searching\\ around for what I want.”\end{tabular}}                                                                                                                             \\ \hline
\rowcolor[HTML]{D9D9D9}
Search                                                                                 & Searching for videos.                                                                                                                                                                       & 36                                                                        & 33\%                                                                                  & \textit{\begin{tabular}[c]{@{}c@{}}“Very efficient and relevant searches.”  —VS.— “Countless \\videos have nothing to do with my latest search request.”\end{tabular}}                                                                                   \\ \hline
Subscriptions                                                                          & Follow specific video creators.                                                                                                                                                             & 35                                                                        & 0\%                                                                                   & \textit{\begin{tabular}[c]{@{}c@{}}“I can choose the content creators I want to follow so that\\ I can limit my time to specific creators I enjoy the most.”\end{tabular}}                                                                                                                                                 \\ \hline
\rowcolor[HTML]{D9D9D9}
Autoplay                                                                               & \begin{tabular}[c]{@{}c@{}}Automatically plays a new video\\ after the current one. Can be\\toggled on/off.\end{tabular}                            & 32                                                                        & 87\%                                                                                  & \textit{\begin{tabular}[c]{@{}c@{}}“I feel like I have little control whenever YouTube takes it\\ upon itself to just play whatever it feels like playing.”\end{tabular}}                                                                                                                                                            \\ \hline
\begin{tabular}[c]{@{}c@{}}Watch history \\\& stats\end{tabular}                         & \begin{tabular}[c]{@{}c@{}}A chronological record of videos\\ watched and time watched stats in\\ YouTube.\end{tabular}                                                                     & 28                                                                        & 7\%                                                                                   & \textit{\begin{tabular}[c]{@{}c@{}}“I am able to view EVERYTHING I do in the app. I\\ can keep an eye if I need to change behavior, what type of\\ videos I watch, everything.“\end{tabular}}                                                                                                                                        \\ \hline
\rowcolor[HTML]{D9D9D9}
Play controls                                                                          & \begin{tabular}[c]{@{}c@{}}Controls to play/pause, seek for-\\ ward/back, etc.\end{tabular}                                                                                                 & 24                                                                        & 12\%                                                                                  & \textit{\begin{tabular}[c]{@{}c@{}}“I can start, pause and stop content streaming easily, at\\ any time.”\end{tabular}}                                                                                                                                                                                                              \\ \hline
Notifications                                                                          & \begin{tabular}[c]{@{}c@{}}System and in-app alerts with new\\ subscription content, recommenda-\\ tions, etc.\end{tabular}                                                                 & 15                                                                        & 53\%                                                                                  & \textit{\begin{tabular}[c]{@{}c@{}}“If I especially like a channel I can know about everything\\ they upload as soon as they do.” —VS.— “Notifications\\ draw me to YouTube and create my schedule for 20-30 \\ minutes. This creates an addiction.”\end{tabular}} \\ \hline
\end{tabular}%
}
\caption{This table shows nine design mechanisms that were mentioned 15 or more times in response to the survey question: \textit{$``$What are 3 things about the mobile app that lead you to feel\textbf{ [most $ \vert $  least]} in control over how you spend your time on YouTube?$"$} Design mechanisms are shown in the order of frequency of mention. The most frequently mentioned mechanism, recommendations, is shown with 3 subcodes. The representative quote(s) column shows one typical response for each design mechanism; both a $``$more in control$"$  and a $``$less in control$"$  quote are shown if the minority opinion on the direction of control was more than 20$\%$ of total responses.}
\label{tab:tab3}
\end{table*}

\noindent \textbf{How Existing Mechanisms Influence Sense of Agency}~\\
The design mechanisms we identified in the YouTube mobile app informed three higher-level themes. First, users experience actions in the app along a \textit{spectrum of consent}. Second, mechanisms for \textit{planning ahead} help them feel more in control. Third, the \textit{accuracy of YouTube algorithms }has mixed consequences for control. The writeup for each theme draws upon examples from our coding of the design mechanisms. \par


\subsubsection{The spectrum of consent.} Participants’ sense of agency depended on whether it felt like they had ‘agreed’ to the actions of the app. Participants gave their active consent through actions such as tapping on a play control: $``$\textit{I’m watching a video that's taken too long of my time, so I can just pause it and come back to it. I feel control there.$"$} Participants could also issue ongoing consent for the app, e.g., by subscribing to a creator: \textit{$``$My subscriptions show me what I asked to see and I can choose what and when I wish to watch each video.$"$} At the other end of the spectrum were mechanisms like autoplay that acted without consent: $``$\textit{It feels weird for the app to start acting before I've told it to do anything.$"$}\par

Non-consent was often felt as a result of (perceived) deception. For example, users disliked ads, but also expected them and indicated their reluctant consent. However, they seemed more upset when the app was unpredictable or violated expectations, as in: \textit{$``$I understand the reason for the ads, but I don't get why some are 5 seconds and you can skip them while others are 60 seconds and you can't.$"$} Other cases where participants felt manipulated included when a \textit{$``$small accidental click$"$ } triggered an ad, when video creators were \textit{$``$not upfront$"$  }about the products they promoted, and when autoplay \textit{$``$automatically$"$} turned on. Participants disliked when the app openly acted against their interests, but expressed stronger sentiments when they felt that the app also misled them about it.

\subsubsection{Planning ahead.} Participants felt more in control when they planned their consumption in advance. Playlists helped participants plan \textit{how much} to watch (e.g., $``$\textit{I can create playlists or queue videos in advance to limit what I watch to a specific list instead of endlessly searching around for what I want$"$}). Participants described the end of a playlist as a \textit{$``$good place to stop$"$,} in contrast to browsing recommendations, which they described as $``$endless.$"$ Watch Later, a default playlist on YouTube, also let participants control \textit{when} and \textit{where} to watch. A guitar teacher described how Watch Later empowered them to save videos on-the-go and watch them later in their music studio. Watch history $\&$ stats also supported planning by providing an awareness that participants could use to adjust their behavior: $``$\textit{I can look at my watch history and see how many videos I have watched today. That puts it into perspective if I should spend time doing something else if I am spending too much time on YouTube.$"$} Several participants described using this awareness in conjunction with the Watch Later playlist: \textit{$``$I am able to put a video in my Watch Later playlist if I think I have spent too much time on YouTube for the day.$"$ }\par

By contrast, sense of agency was diminished by mechanisms that prompted and pressured participants with suggestions that were hard to decline. Autoplay and recommendations frequently led to this, as in \textit{$``$I often spend more time than I meant to because there is a good related video that seems worth watching so ya know, ‘Just one more’ which becomes a couple hours.$"$ } The Watch Later playlist again served as a safety valve in \textit{‘}just one more\textit{’} situations: \textit{$``$Watch Later means I don't feel pressured into watching a recommended video from autoplay right when I see it.$"$}\par

Notifications sometimes supported planning and sometimes not. For example, they put participants on the spot:\textit{ $``$Based on my viewing history, the app will push me new content and I may not have the fortitude to not click to view.$"$  }However, notifications also helped participants plan when to check the app by reducing their fear of missing out: $``$\textit{With notifications I will know exactly when a new video goes up that I may be interested in watching. This way I am not randomly checking for uploads and spending extra time searching and browsing.$"$  }This may explain why notifications were split between $``$more in control$"$  and $``$less in control$"$  responses (47$\%$  vs. 53$\%$ ).

\subsubsection{The accuracy of algorithms has mixed consequences for control.} Irrelevant recommendations, i.e., those that were repetitive or unrelated to personal interests, universally undermined sense of agency:\textit{ $``$Seeing 'recommended' videos that have nothing to do with my viewing history leads to unwanted scrolling and possibly unwanted content.$"$} Similarly, irrelevant search results undermined control because they forced participants to keep scrolling for what they wanted, e.g., \textit{$``$I use specific search terms, but I still have to scan past a lot of vaguely or even unrelated stuff to find what I want.$"$  }\par
For relevant recommendations, participants’ control responses were divided nearly 50-50. In contrast to irrelevant recommendations, relevant ones supported control with their personalization (e.g., \textit{$``$It has some very good algorithms that know what I like, when I want it$"$ }) or with suggestions that reached just beyond the users’ comfort zone (e.g., \textit{$``$I can expand my tastes based on my own preference$"$ }). However, relevant recommendations sometimes undermined control by being \textit{too engaging}, i.e., recommending videos that users watch, but that are unplanned and later regretted. This was captured in participants’ use of terms like the $``$\textit{wormhole$"$ } (two mentions) and $``$\textit{rabbit hole$"$ } (five mentions), as in $``$\textit{The way that videos get promoted to my home page and have appealing thumbnails--I end up clicking on them and wonder how I got to this place and why I am watching this video. I ended up going down the rabbit hole and watching the video and then others like it and so on.$"$  }Some of these recommendations were described as \textit{$``$clickbait$"$  }(six mentions) that misled with content that did not meet expectations and sometimes also violated participants’ consent (e.g., by showing \textit{$``$inappropriate content$"$}). More often though, participants seemed to like the content, but felt that it was too much (e.g., $``$\textit{At times there is no escape when I become interested in documentary after documentary$"$}) or not the right time (e.g., $``$\textit{Some of the church videos are addicting and I keep watching them at night$"$).}\par
Given their mixed experiences with recommendations, participants expressed frustration with the customization settings at their disposal (or lack thereof). Participants lacked the ability to customize the location, quantity, and content of recommendations. Having recommendations on almost every screen led to a loss of control: $``$\textit{It seems like there are video recommendations everywhere. They} \textit{are obviously in my home feed; they are in the explore menu; and they are under and beside and within other videos. It often takes me down the rabbit hole.$"$ } Up next recommendations that appear below the current video (and autoplay after it finishes) were specifically mentioned seven times. The \textit{$``$endless$"$} quantity of recommendations also made it hard to stop watching. Finally, participants also wanted to control \textit{what} content is recommended, particularly when recommended content did not match their aspirations: \textit{$``$There are cases in a particular day where I just want to watch cat videos. But I do not want my entire screen to recommend cat videos.$"$} Participants wanted to customize the content of recommendations more directly than just by generating a watch history: \textit{$``$The only thing you can do to control the algorithm is to watch videos. But you get no say how it'll recommend new ones.$"$}\par

A minority of responses described recommendation settings that \textit{do} support sense of agency. For instance, three participants appreciated how the settings menu ($ \threedotsopen $) allows them to mark $``$Not interested$"$ on specific videos, e.g., $``$\textit{When I'm tempted but know a video is not educational I can hide it.$"$} In this case, the user \textit{is} in fact interested in the sense that the video recommendation arouses their curiosity and attention. However, they must paradoxically mark it as ``Not interested'' in order to tell the interface to stop showing videos of this kind because they conflict with their longer-term goals. YouTube’s settings also allow participants to delete videos from their watch history--which stops them from being used in personalized recommendations--but only one participant mentioned this feature. The vast majority of participants seemed either unaware of YouTube’s existing customization settings for recommendations or found them inadequate. \par

\section{Study 2: Co-design with YouTube users}
Study 1 identified existing mechanisms in the YouTube mobile app that influence user sense of agency (\textbf{RQ1}). In Study 2, we sought to understand how \textit{changes} to these design mechanisms might influence sense of agency (\textbf{RQ2}). We conducted 13 study sessions with individual YouTube users that included two co-design activities: 1) sketching participant-generated changes; and 2) evaluating researcher-generated changes that were based on the results of Study 1. Consistent with a research-through-design approach \cite{Zimmerman2014-pu}, the aim of these activities was not to converge upon a single solution but rather to generate knowledge, i.e., what to design for a sense of agency.\par

\subsection{Preparatory Design Work}
In preparation for the evaluation co-design activity, five of the authors (KL, HZ, JVL, JC, KF), all advanced-degree students in a technology design program, created mockups of changes to mechanisms in the YouTube mobile app that we expected to impact sense of agency. To generate a wide range of possible changes, we started with a design brainstorm that generated 67 different ideas, e.g., creating a ‘How-to mode’ for viewing only educational content, reducing video playback speed to 50$\%$  after a daily time limit is exceeded, or making Watch Later the default action for recommendations. Ideas were reviewed as a group and favorites could be ‘claimed’ by one author who further refined it. This generated a total of 33 different sketches. We presented, discussed, and then scored these sketches according to three criteria: expected impact on sense of agency (based on the results of Study 1), novelty relative to existing digital wellbeing tools, and feasibility of implementation.\footnote{Feasibility was a criterion to focus on designs that a third-party mobile developer could build using public APIs, an intention we have for our future work.} Expected effect on sense of agency was weighted twice in our scoring.\par

We created mockups for the seven sketches with the highest average scores. We wanted participants to evaluate a variety of potential changes to each mechanism, so we created three versions of each mockup: low, medium, and high-control. For example, the recommendations mechanism in the YouTube app was redesigned to change the number of recommendations shown on the homepage, with the low-control version showing unlimited recommendations, the medium-control version showing only three recommendations with a button to $``$show more,$"$ and the high-control version not showing any recommendations (see images in \textbf{Table \ref{tab:tab4}}). To focus on RQ2, our results and analysis here address only the four mockups (see \textbf{Table \ref{tab:tab5}}) that directly change one of the existing internal mechanisms in YouTube that we identified in Study 1. The other three mockups we tested—activity-goal setting, time-goal setting, and a timer—are more external mechanisms that might apply equally well to other apps. However, we decided to focus this paper on the unique potential of internal mechanisms.

We note that although our research focuses at the level of ‘design mechanisms,’ the details of these designs matter. For instance, although the recommendations in the current version of YouTube seemed to reduce sense of agency in most of the Study 1 responses, a different implementation of ‘recommendations’ might produce different effects. This is true of our mockups too: in our search redesign we showed a task-oriented example query \textit{($``$How to cook a turkey$"$)}, whereas a leisure-oriented example query (e.g., \textit{$``$Funny cat videos$"$}) could have led to different results. We include descriptions of the most relevant details of each of these design mechanisms in the body of the paper, screenshots of their current implementation in the YouTube mobile app in \textbf{Appendix I}, and images of \textit{all} of our mockups in \textbf{Appendix II}.\par

\begin{table*}
\tiny
\centering
\resizebox{\textwidth}{!}{%
\begin{tabular}{c|c|c}

\\ 
\begin{tabular}[c]{@{}c@{}}Low-control version: \\ Unlimited recommendations\end{tabular} & \begin{tabular}[c]{@{}c@{}}Medium-control version: \\ Click-to-show-more-recommendations \end{tabular} & \begin{tabular}[c]{@{}c@{}}High-control version: \\ No recommendations\end{tabular} \\ 
\raisebox{-.010\height}{ \includegraphics[scale=0.10]{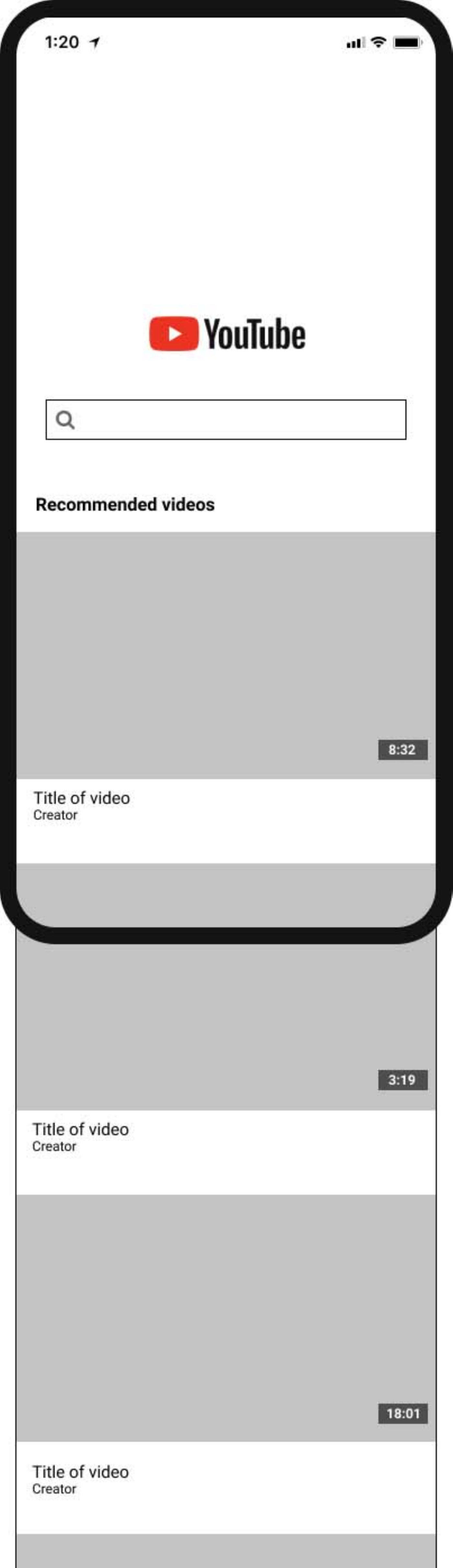}}
&           
\includegraphics[scale=0.10]{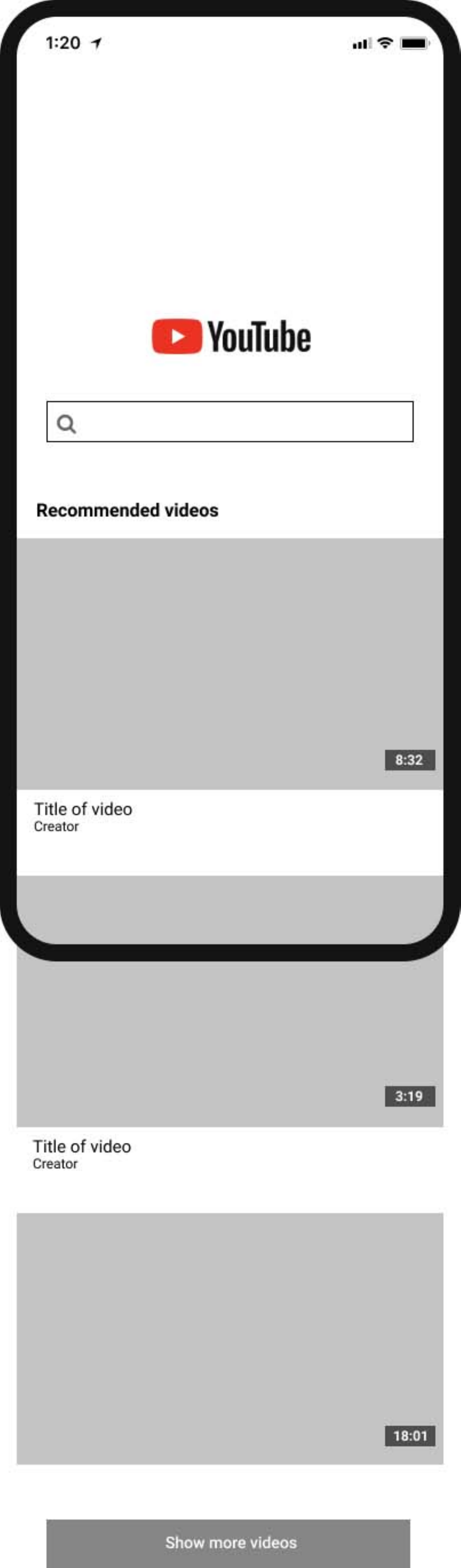}
&         
\raisebox{.62\height}{\includegraphics[scale=0.10]{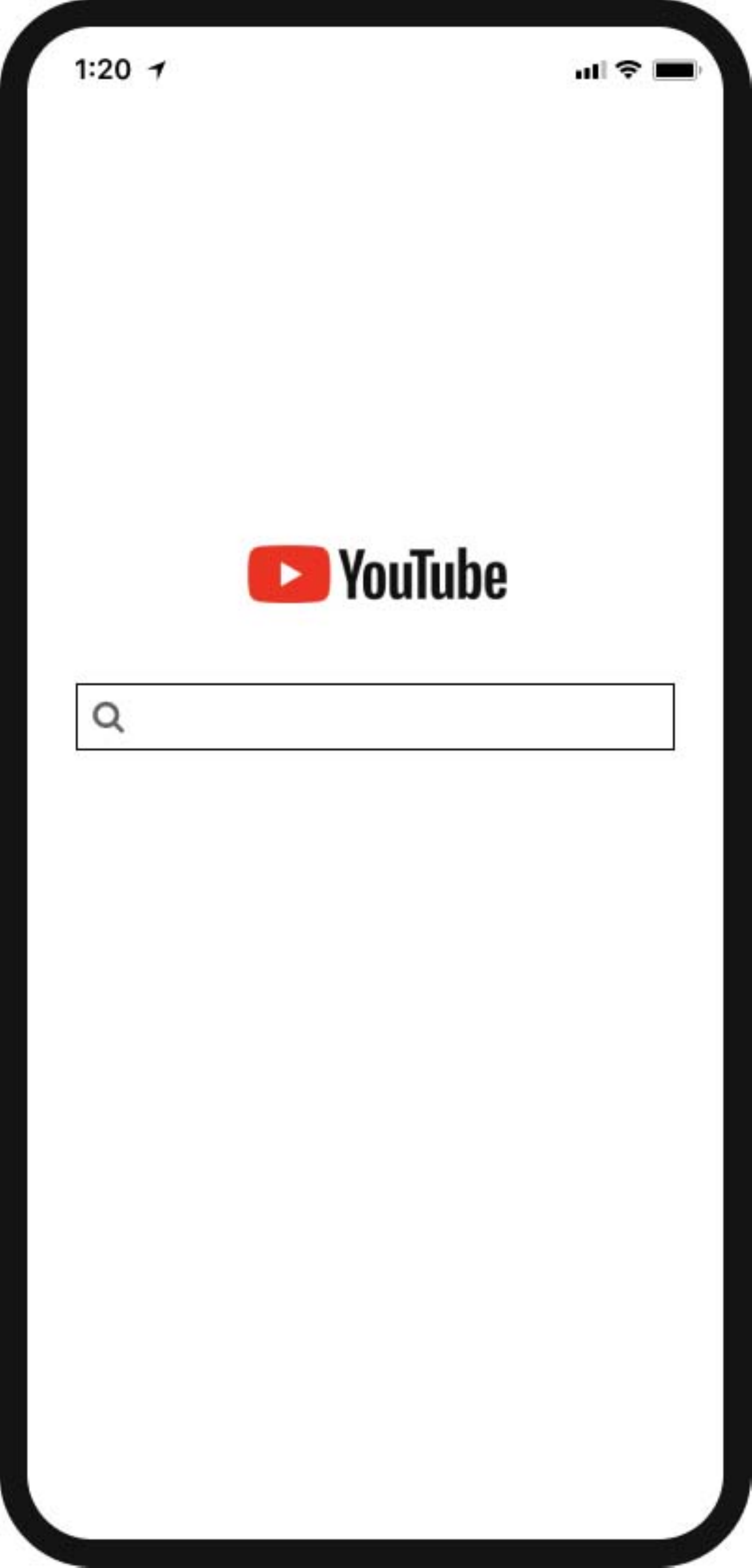}}
\\
\end{tabular}%
}
\caption{Mockups of the redesign of the recommendations mechanism. We created three versions of the mockup that we expected to offer different levels of control. These 3 versions of each redesign were evaluated by participants in the co-design evaluation activity.}
\label{tab:tab4}
\end{table*}

\begin{table*}
\centering
\resizebox{\textwidth}{!}{%
\begin{tabular}{|c|c|c|c|c|c|c|}
\hline
\textbf{\begin{tabular}[c]{@{}c@{}}Redesigned\\ mechanism\end{tabular}} & \textbf{Dimension of change}                                                                                                                     & \textbf{\begin{tabular}[c]{@{}c@{}}Low-control \\ version\end{tabular}}                           & \textbf{\begin{tabular}[c]{@{}c@{}}Medium-control\\ version\end{tabular}}                                     & \textbf{\begin{tabular}[c]{@{}c@{}}High-control \\ version\end{tabular}}  & \textbf{\begin{tabular}[c]{@{}c@{}}Related experience for users (as de-\\ scribed by Study 1 participants)\end{tabular}}                    & \textbf{\begin{tabular}[c]{@{}c@{}}Comparison to\\ current version of\\ YouTube mobile app\end{tabular}} \\ \hline
Recommendations                                                         & \begin{tabular}[c]{@{}c@{}}Number of video recom-\\ mendations on home screen\end{tabular}                                                       & \begin{tabular}[c]{@{}c@{}}Unlimited recom-\\ mendations\end{tabular}                              & \begin{tabular}[c]{@{}c@{}}Shows 3 recommen-\\ dations, then a click-\\ to-show-more button\end{tabular} & \begin{tabular}[c]{@{}c@{}}No \\ recommendations\end{tabular}             & \begin{tabular}[c]{@{}c@{}}Endless recommendations often \\ undermine sense of agency\end{tabular}                                       & \begin{tabular}[c]{@{}c@{}}Similar to low-control\\ version\end{tabular}                                 \\ \hline
Playlists                                                               & \begin{tabular}[c]{@{}c@{}}Prominence of button to\\ save a video to the Watch\\ later playlist\end{tabular}                                     & \begin{tabular}[c]{@{}c@{}}No Watch Later \\ button\end{tabular}                                 & \begin{tabular}[c]{@{}c@{}}Small Watch later\\ button\end{tabular}                                            & \begin{tabular}[c]{@{}c@{}}Large Watch Later\\ button\end{tabular}          & \begin{tabular}[c]{@{}c@{}}Watch Later playlist lets users plan\\ ahead, reduces pressure to watch now\end{tabular}                        & \begin{tabular}[c]{@{}c@{}}Similar to medium-\\ control version\end{tabular}                             \\ \hline
Search                                                                  & \begin{tabular}[c]{@{}c@{}}The degree to which search\\ prioritizes fun vs. relevant\\ results (see \textbf{Appendix II} for\\ more details)\end{tabular} & \begin{tabular}[c]{@{}c@{}}Prioritize “fun” re-\\ sults (intended to\\ be too engaging)\end{tabular} & \begin{tabular}[c]{@{}c@{}}User can toggle \\ between “fun” \&\\ “relevant” results\end{tabular}                 & \begin{tabular}[c]{@{}c@{}}Prioritize “relevant” \\ results\end{tabular} & \begin{tabular}[c]{@{}c@{}}Sometimes recommendations and \\ search results that are too engaging\\ undermine sense of agency\end{tabular} & \begin{tabular}[c]{@{}c@{}}Similar to medium-\\ control version\end{tabular}                             \\ \hline
Autoplay                                                                & \begin{tabular}[c]{@{}c@{}}The degree of user consent\\ required to play the next\\ video recommendation\end{tabular}                            & \begin{tabular}[c]{@{}c@{}}Autoplay the next\\ recommendation\end{tabular}                     & \begin{tabular}[c]{@{}c@{}}Show the next recom-\\ mendation\end{tabular}
   & \begin{tabular}[c]{@{}c@{}}No next \\ recommendation\end{tabular} 
   & \begin{tabular}[c]{@{}c@{}}Autoplaying videos without consent un-\\ dermines sense of agency\end{tabular}                                 & \begin{tabular}[c]{@{}c@{}}Similar to low-control\\ version\end{tabular}                                 \\ \hline
\end{tabular}%
}
\caption{This table describes our redesigns of 4 existing mechanisms in the YouTube app. We created three versions of each mockup that we expected to provide different levels of control to the user: low, medium, and high. \textbf{Appendix II} describes more details about the search redesign and the three additional mockups we created, which we do not report on here.}
\label{tab:tab5}
\end{table*}

\subsection{Participants}
\subsubsection{Recruitment.} We recruited YouTube users in Seattle via email lists and social media channels to \textit{$``$Help us understand how people spend their time in the YouTube mobile app.$"$ } We did not initially set inclusion criteria for participation (beyond adult YouTube users) as we viewed our co-design activities as exploratory. However, after our initial sessions proved insightful for our team of design researchers, we sent a follow-up survey to participants that asked about demographics and YouTube use. Participants were compensated with a $\$$30 voucher.\par

\subsubsection{Demographics and YouTube use.} 13 YouTube users (7 women, 6 men) participated in our sessions. The median age was 29 (range: 18-36). Participants reported using YouTube a median of 52 minutes per day (range: 27-70), again based on checking their time watched statistics in the YouTube mobile app. For reference, this amount of time is slightly lower than the average of signed-in YouTube users (60 minutes) \cite{Matney2017-ac} and considerably lower than the median of participants in Study 1 (101 minutes). \newline

\subsection{Procedures}
Sessions included an initial think-aloud demonstration of their current YouTube use, followed by sketching and evaluation co-design activities. The median length of a session was 73 minutes (range: 57-105 minutes). \par

\subsubsection{Think-aloud Demonstrations with YouTube App.} In a modified version of a think-aloud-protocol \cite{Jaaskelainen2010-ts}, the participant opened YouTube on their smartphone and talked us through a typical engagement cycle (how they start and stop use) \cite{Tran2019-lv}. Next, they showed and talked us through the mechanisms that made them feel most and least in control of how they spend their time on YouTube. \par

\subsubsection{Co-design Activity 1: Sketching.} 
\begin{figure}
    \centering
    \includegraphics[scale=0.124]{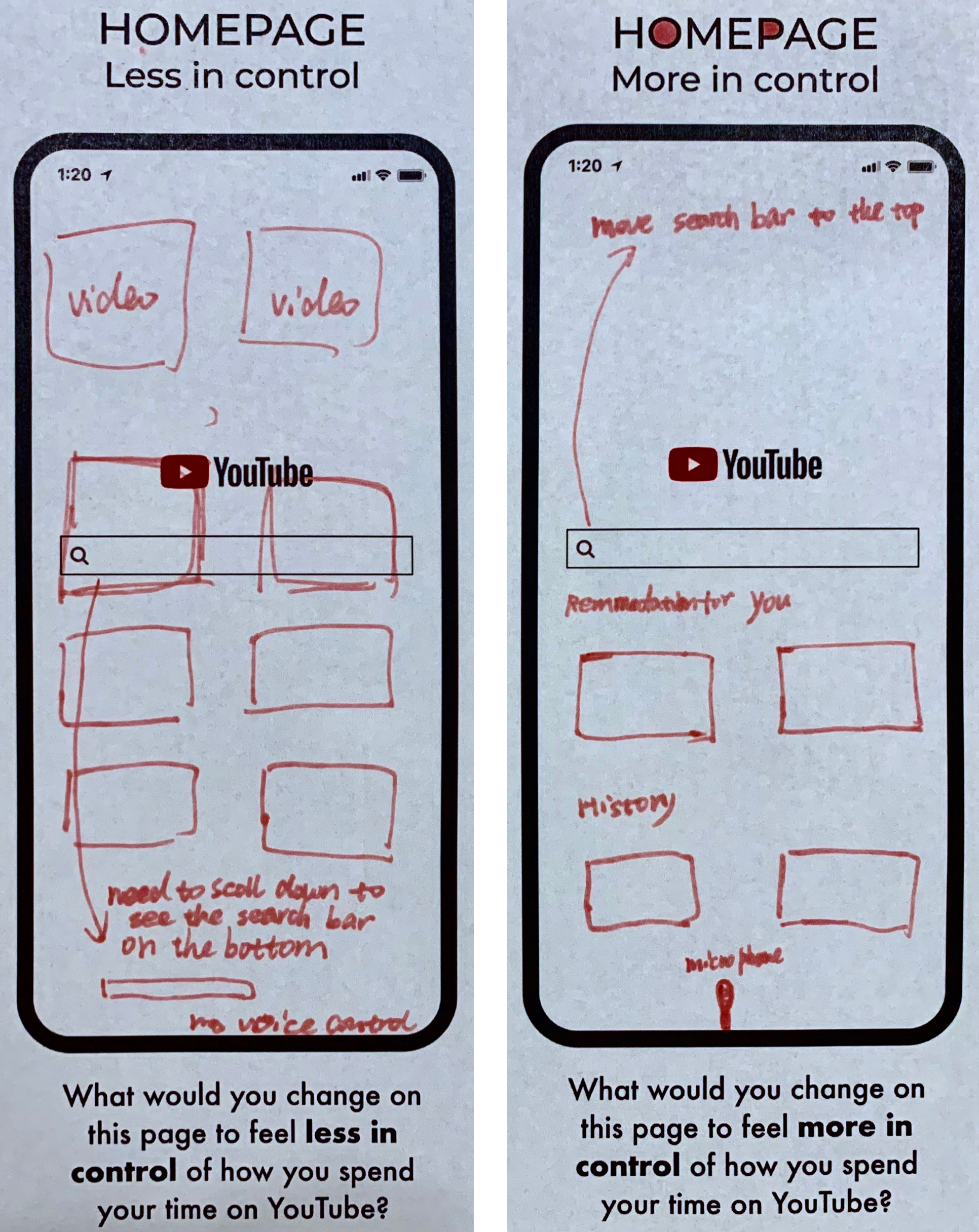}
    \caption{Sketches of the home screen of the YouTube mobile app. The participant (P11) explained that in the  $``$more in control$"$  version, recommendations are based on topics chosen by the user. In the $``$less in control$"$  version, the user needs to scroll through recommendations to see the search bar at the bottom of the screen.}
    \label{fig:fig3}
\end{figure}

To elicit participant-generated ideas, we asked participants to sketch over paper mockups of three key screens: home, search, and video player (see \textbf{Figure \ref{fig:fig3}}). Each screen represented a minimal version of a video app without recommendations, rather than a direct copy of the current YouTube interface. We chose this minimal version to encourage participants to generate new ideas, rather than to evaluate the existing interface (which we did in Study 1). Participants were handed a pen and a copy of one mockup (e.g., the home screen) and were asked, $``$\textit{What would you change on this page to feel more in control of how you spend your time on YouTube?$"$ } They then received a second copy of the same mockup and were asked to sketch changes that would make them feel \textit{$``$less in control.$"$ } Each participant created a total of six sketches (two versions of three different screens). As they sketched, participants were asked to explain their thinking \cite{Schrage1996-hh}.\par

\subsubsection{Co-design Activity 2: Evaluation.} To receive feedback on our changes from YouTube users, we asked participants to evaluate our mockups of the redesigned mechanisms in the YouTube mobile app (see \textbf{{Table \ref{tab:tab5}}}). For each mockup, the three different versions were placed in front of the participant in a random order, they reviewed for about one minute, and then asked any questions they had. We did not tell participants which one was the low, medium, or high-control version. The participant was then asked to rank the three versions in order from the one they would least prefer to use to the one they would most prefer, and explain why.\par

\subsection{Codebook Thematic Analysis}~\\
We used codebook thematic analysis to analyze the data \cite{Braun2018-th,Braun2006-ma}, wherein we generated themes that are more interpretive than just a summary of all of the data, but less interpretive than in reflexive thematic analysis where the researcher’s subject position plays a central role in the analysis \cite{Braun2019-af}. After each co-design session, the researcher leading the session completed a debriefing form with their top three takeaways and shared participant sketches with the rest of the research team. We held weekly meetings to discuss these data and discuss initial ideas. After finishing data collection, all co-design sessions were transcribed. To further familiarize ourselves with the data, three of the authors read the transcripts and again reviewed the sketches. We next independently coded the data using a web app for collaborative coding \cite{Sillito_undated-xq} to generate our set of initial codes. After reviewing this first pass of coding together, we refined and consolidated codes and generated initial themes. Our final set of codes included: user freedom of choice, situational features affecting control, design mechanisms for control, setting clear expectations for the user, and triggers to stop, each of which had further subcodes. We applied our codes to all transcripts and sketches and reviewed the results to create our final themes. For each theme, we extracted vivid exhibits \cite{Bannon1994-iy}, which we used to write analytical memos.\par

\subsection{Results and Analysis}
We generated two themes about how participants expected changes to the design mechanisms of YouTube would affect their sense of agency. First, participants wanted design mechanisms that provided \textit{more control when they had an intention in mind} as opposed to when they just wanted to explore. Second, participants envisioned and wanted mechanisms for \textit{active and informed choices} to increase control. \par

\subsubsection{Specific intentions call for more control.} 
When individual participants reviewed the different versions of their own sketches and our mockups, they were often conflicted about how much control they preferred. It depended upon the situation. When they had a specific intention or goal for their YouTube visit (e.g., to cook a recipe), they wanted design mechanisms that provided greater control. When they had a non-specific intention such as relaxing, they preferred design mechanisms that turned control over to YouTube. \par

For participants, specific intentions varied from watching a video of a favorite dance, to the latest basketball highlight, to a tutorial on solving a Rubik's Cube. When they had such a specific intention in mind, they wanted greater control than YouTube currently gives them. P4 removed recommendations from their sketch, explaining: \textit{$``$If I have a specific goal, I know what I want, I don't need recommendations to guide my search, I just want to be in control of my search.$"$} P2 evaluated our redesign of the search mechanism that emphasized results with higher entertainment value by saying, \textit{$``$I'm probably going to click on it because it's cute and I'm just going to waste so much time. So it's going to make me feel totally out of control of what I actually wanted to come here for.$"$} In these cases, participants wanted stronger control mechanisms so that the app would not hijack their specific intention.\par

Sometimes participants held intentions with a moderate level of specificity, in which case participants wanted to retain some control but also delegate some to YouTube. Often these intentions were topical, as in when P11 wanted to be able to use the app in an \textit{$``$active way$"$} to search and browse videos about programming, but not in a \textit{$``$passive way$"$} to follow just any recommendation. Sometimes, these intentions were temporal, such as when working or studying, participants preferred a version of YouTube that helps them watch a moderate number of videos without making them \textit{$``$fall down a rabbit hole of similar related stuff$"$} (P13). To address these cases, participants sketched both changes to internal mechanisms that were specific to YouTube (e.g., limits on the number of recommended videos) and also more external mechanisms that might apply across a variety of social media apps (e.g., time reminders). \par

By contrast, when participants had only a non-specific intention (e.g., to unwind or explore), they wanted YouTube to lead the way. Our redesigns of the recommendations mechanism showed either unlimited, limited, or no video recommendations, to which P2 responded \textit{$``$If I came here for a specific reason, like my goal is to learn how-to do something, then I prefer this one without recommendations. However, if I just want to watch something that gets my mind off things, I prefer the one where I can choose to show more recommendations.$"$} At times when participants just wanted to be entertained, designing for greater control could actually get in the way. P13 shared, \textit{$``$If you're not giving me recommendations, and if you're making me search, then I'm not in control. Or, I'm in control, but the problem is I’m spending more time. There’s no point.$"$ } \par

\subsubsection{Active and informed choices.} The\ Study 1 theme $``$Spectrum of consent$"$  addressed whether the user had 'agreed' to an action taken by the app (e.g., autoplaying the next video). To support control, Study 2 participants envisioned more \textit{active} choices, where the user felt like they were the one to initiate the action. As a step in this direction, P1 described a home screen that presented, \textit{$``$Six categories we think you're most interested in, and then you're at least making the active choice, ‘I want to watch some interviews right now.’$"$  }In this design, the app’s algorithm would recommend a set of personalized topics, but the user would be the one to choose between them. A still more active choice was when the user was the one to generate the set of choices in the first place, as in P7’s sketch: $``$\textit{There aren't a billion recommendations on the home screen. It's just a search bar. You go straight to what you want to watch, you watch it, and then you're done.$"$  }Participants described search as a paragon of user-led choice, and many foregrounded the search option in their sketches to increase control and hid it in ones to decrease control (see \textbf{Figure \ref{fig:fig3}}).\par

Many sketches also supported more \textit{informed} choices. These designs made it easier for users to know what to expect from a video by surfacing metadata like view count, user ratings, and descriptions. Five participants proposed novel metadata, such as an ‘activity time’ filter that would sort how-to videos by the time it takes to perform the activity they teach, e.g., cook a recipe (P12). Another suggested expert ratings as an indicator of quality (P5). Conversely, in sketches to undermine control, it was common to remove video metadata. P12 likened this to the experience at Costco, a supermarket chain that deliberately shows no signs in its stores \cite{Npr2015-ps}: \textit{$``$If you want to go find cookies, they won't actually show you where the cookies are so you literally have to go through every single aisle. You have to go find it.$"$  }\par

More choice alone did \textit{not} lead to more control. In sketches of designs to undermine control, participants covered every corner of the home screen with video recommendations that scrolled infinitely (P11) and in every direction (P5). P13 described, \textit{$``$If they didn't have [recommended videos], it would be a lot harder to follow these different rabbit holes. I imagine that I would have to intentionally seek out another video, so I wouldn't feel sucked in as much.$"$ } Recommendations prompted a passive form of choice, in which users reacted to the app’s infinite scroll of suggestions, rather than making active choices on their own terms. \par

\section{Overall Discussion}
Together, our two studies identify design mechanisms that influence sense of agency in the YouTube mobile app and how they might be changed to increase it. In Study 1, participants reported that, in the current app, recommendations, ads, and autoplay mostly led them to feel less in control, whereas playlists, search, subscriptions, play controls, and watch history \& stats mostly made them feel more in control. Across all existing mechanisms, participants felt less in control when the app took actions of its own without their consent (e.g., autoplaying a new video recommendation). Recommendations were of special concern and participants expressed frustration at their inability to customize their location, quantity, and content. In contrast, by helping participants plan ahead for even just a short while, existing mechanisms like playlists and watch stats made participants feel more in control.

When participants envisioned and evaluated changes in Study 2, they wanted more opportunities to make active choices, rather than respond to a set of choices proposed by the app. This preference was stronger when they had a specific intention in mind (e.g., to watch a certain video or topic), whereas when their intention was more general (e.g., to pass the time) they favored turning control over to YouTube. \par

We expect that our findings on how design mechanisms influence sense of agency on YouTube are most likely to generalize to other social media and media apps where users (a) report feeling out of control at times (e.g., Facebook \cite{Marino2018-qb}); and (b) use the app for both specific and non-specific intentions (e.g., Pinterest \cite{Cheng2019-me}). We first discuss our findings mostly with respect to our test case of YouTube, before considering implications for digital wellbeing more broadly.

\subsection{Rethinking What ‘Relevance' Means for Recommendations}
Recommendations were mentioned by participants as undermining sense of agency far more times than any other design mechanism in the YouTube mobile app, suggesting that recommender systems \cite{Resnick1997-rd} should be of central concern to digital wellbeing designers. However, they led to a reduced sense of agency via two very different routes: irrelevance and relevance.\par

First, recommendations were sometimes irrelevant, showing videos that participants were simply not interested in. However, due to rapid advances in artificial intelligence and recommender systems like YouTube specifically (e.g., \cite{Covington2016-nf}), one might expect recommendations in social media apps to become more and more relevant in the coming years. \par

Second, recommendations were sometimes too ‘relevant,’ which presents a more vexing problem from a digital wellbeing perspective. For example, participants reported that they sometimes saw \textit{too many} interesting recommendations (e.g., for documentaries or for church videos late at night), which made them feel a loss of control. In this case, YouTube's algorithm is arguably \textit{too good} at a local optimization problem: \textit{Out of millions of videos, which one is the user most likely to watch?} But it misses a more global optimization problem: \textit{Out of many possible actions, which one does the user most want to take?} In these cases, recommendations appealed to a users' impulse or short-term desire to watch more videos, but conflicted with their long-term goals, creating a self-control dilemma for the user \cite{Lyngs2019-mw, Duckworth2016-mq}. 

Our findings call for rethinking what `relevance' means for recommendations in the context of digital wellbeing. Prior research on recommender systems has argued that \textit{``being accurate is not enough,''} as a fixation on accuracy can lead designers to ignore important facets of user experience like serendipity \cite[p.1]{McNee2006-ji}. For participants in our study, sense of agency was clearly a neglected facet of user experience, as YouTube's recommendations led them to actions (i.e., watching more videos) they did not feel in control of. To be clear, this does not mean that Google or others should try to create an `algorithm for life’ that recommends between watching another video, writing a term paper, and going to sleep. 

However, it does suggest that recommender systems could first start with the global problem of \textit{when} to show recommendations, before moving on to the local problem of \textit{which} items to recommend. For example, a decision \textit{not} to show recommendations might be informed by the time of day (e.g., 2am is too late), screentime preferences (e.g., when the user has already exceeded their goal of 30-minutes per day on entertainment apps), or explicit user preferences (e.g., only show three recommendations unless I click-to-show-more). In HCI research, sometimes the implication of a user needs assessment is \textit{not} to design technology, as a new technology might not be appropriate in the context of the larger situation \cite{Baumer2011-ho}. Similarly, for recommender systems, our findings suggest that sometimes the implication is \textit{not} to recommend. Prior work has addressed how a system can display the level of confidence it has in its recommendations to the user \cite{McNee2003-ni}, but this should be preceded by a more fundamental question of whether or not to show recommendations in the first place.\par

Whereas both of the studies in this work elicit user preferences (``what users say''), the dominant paradigm of recommender systems today, including YouTube, is behaviorism: recommendations largely neglect explicit preferences and instead rely on behavior traces (``what users do'') \cite{Ekstrand2016-qy}. The present bias effect \cite{ODonoghue2015-ue} predicts that \textit{actual behavior} will favor the choice that offers immediate rewards at the expense of long-term goals. In this way, recommender systems reinforce the sometimes problematic behavior of the current self rather than helping people realize their `aspirational self' that reflects long-term goals \cite{Ekstrand2016-qy,Lyngs2018-qi}.  

Participants also wanted to customize the \textit{content} of recommendations, e.g., $``$\textit{I do not want my entire screen to recommend cat videos.$"$} Today, the dominant paradigm of recommender systems, including YouTube, is behaviorism: recommendations rely on behavior traces (``what users do'') and largely neglect explicit preferences (``what users say''). In this way, recommender systems reinforce the sometimes problematic behavior of the current self rather than helping people realize their `aspirational self' that reflects long-term goals \cite{Ekstrand2016-qy,Lyngs2018-qi}. Designers might address this by making it easier for users to (a) explicitly state preferences for topics they would like to see or not see; (b) explicitly rate recommendations (e.g., show me more like this one); (c) edit their viewing history to influence future recommendations (e.g., delete all cat videos); or (d) select an algorithmic personae to curate their recommendations (e.g., $``$The Diplomat,$"$  who brings news videos from the other side) \cite[p.72]{Harambam2019-qu}. The current YouTube app offers limited support for these first three features (e.g., users can select from among topics for recommendations on the home page of the app), but participants in our study seemed mostly either unaware of these customization settings or found them to be inadequate.  \par

To summarize, we encourage designers of recommender systems to think beyond just optimizing for the item that is most likely to be clicked, watched, or liked. This includes considering \textit{when} to show recommendations in the first place. It also means exploring how recommendations can support user aspirations rather than just reinforce current behaviors, which requires better measures of long-term preferences. Designers and researchers should continue to explore how to personalize recommendations to satisfy these broader user needs, or provide customization options that put users in control - at least to the extent they want. \par

\subsection{Designing to Support Microplanning}
Behavior change researchers have long known that plans can help bridge the gap between intentions and behavior. In this work, plans are usually crafted in advance through careful deliberation and guide behavior for some time into the future \cite{Agapie2020-ne}. For example, a screentime tool in this mold might ask the user to review and reflect upon their past usage data and develop a plan for their use over the next month. Participants in our study also ‘planned’, but they did so in a more ad hoc manner. For example, they queued videos in advance to limit what they watched during a single session or glanced at their Time watched statistics to know whether to watch another video or add it to their Watch Later playlist.\par

These types of actions might be called ‘microplanning,’ making lightweight plans that guide behavior for a short time, usually just a single session of use. Our naming takes inspiration from Cox et al.’s coining of the term ‘microboundary’ to describe \textit{$``$a small obstacle prior to an interaction that prevents us rushing from one context to another},\textit{$"$} which serves as a ‘micro’ version of a commitment device that prevents the user from \textit{$``$acting hastily and regretting it later$"$} \cite{Cox2016-ts}. ‘Microboundary’ has helped center an important concept from behavioral economics, commitment devices that restrict future choices to reflect long-term goals \cite{Bryan2010-wt,Schelling1984-zo}, in the research and development of digital wellbeing tools, e.g., \cite{Kim2019-cd,Kim2019-pq,Lyngs2019-mw,Pinder2018-uk}. \par

Similarly, we hope that the concept of ‘microplans’ encourages the use of behavior planning knowledge in the design of digital wellbeing tools. For example, this literature finds that plans are more likely to succeed if they specify where, when, and how a behavior will be enacted \cite{Gollwitzer2006-ua}. A microplan might incorporate just the \textit{where} part, and be supported by a video playlist that is tied to a specific location, e.g., song tutorials \textit{for my guitar studio}. Triggers are also a key component of effective plans \cite{Fogg2009-sy}, so in this case the playlist might be the primary recommendation in the app anytime the user is within 50 meters of the studio. In another example, Hiniker et al. adapted an evidence-based Plan-Do-Review sequence \cite{Felner1988-yr} for an app that asked children to plan out their video-watching, finding that it helped them transition to their next activity with ease \cite{Hiniker2017-sx}. In the domain of impulse buying \cite{Moser2019-vp}, an e-commerce site (or third-party extension) might foreground ‘shopping list’ tools to support intentional buying.\par

\subsection{Different Levels of Control for Ritualized and Instrumental Use} 
In Study 2, participants suggested ways that the YouTube mobile app might be redesigned to increase sense of agency (e.g., by reducing the number of recommendations it displays). However, such changes might lead to adverse effects as there were also times when participants preferred low-control features. Although HCI design guidelines advise supporting user sense of agency \cite{Nielsen1994-ss,Shneiderman2004-yh}, we should not assume that a greater sense of agency is always desirable.\par

Specifically, participants preferred higher-control mechanisms when they had a specific intention in mind and lower-control ones when they had a non-specific intention. This finding broadly aligns with two types of viewing that have been identified in uses and gratifications research on television use \cite{Rubin1984-so}: (1) ritualized use, open-ended use to gratify diversionary needs; and (2) instrumental use, goal-directed use to gratify informational needs. On this view, the current version of the YouTube app appears to offer good support for ritualized use, but poor support for instrumental use, as participants often felt that their specific intentions were hijacked by its autoplay and endless recommendations.\par

How might a single app support sense of agency for both ritualized \textit{and} instrumental use? One approach is a customizable interface that lets the user switch between low and high levels of control. This can be done at the app-level, e.g., switching between an Explore Mode and a Focus Mode. Or it can be done at a mechanism-level, e.g., YouTube currently offers an on/off toggle for autoplay, but does not provide any way to toggle recommendations, which were the mechanism most frequently mentioned as leading to a loss of control in Study 1. This approach may be particularly suitable for power users, as prior research indicates that power users prefer interfaces that are customizable (user-tailored) by a toggle, whereas non-power users prefer ones that are personalized (system-tailored) for them \cite{Sundar2010-oh}. \par

A second approach then is an interface that is personalized for the user based on a prediction model. Recent work has found that classifiers can be trained to predict these types of media use with high confidence, e.g., for Pinterest \cite{Cheng2017-ge} and smartphone use \cite{Hiniker2016-ef}. For example, if YouTube expects that the user is visiting for ritualistic use, it could remain as is, or even go further to take control as in its Leanback mode for \textit{$``$effortless viewing$"$} that autoplays a never-ending stream of high-definition recommendations \cite{Google_undated-yh}. Both our own findings on autoplay and previous work suggest that such a high level of automation would reduce sense of agency \cite{Berberian2012-gt}, but it may still be the interface that the user prefers in this situation. Conversely, if YouTube has high confidence that the user is visiting for instrumental use, it could present a search-only interface and hide all recommendations. Finally, if it has low confidence in its prediction, it could present a middle-ground interface that shows limited recommendations, or it might err on the side of caution and lead with a search-first interface in case the user has an intention to express. \par

\subsection{Towards a Language of Attention Capture Dark Patterns}
Our findings address \textit{what} and \textit{when} to design to increase sense of agency. However, in the attention economy, what might motivate key stakeholders to support such designs? One step is for the design community to develop a common language of attention capture dark patterns that recognizes designs that lead to attentional harms. 

Developing such a lingua franca of attention capture design patterns could be integrated into design education \cite{Gray2018-zz}, influence designer thinking, and reputations, as is done by the name-and-shame campaign of the darkpatterns.org website \cite{Brignull_undated-pv}. At the company level, it could help inspire products that are mindful of the user’s sense of agency. For example, in spite of the incentives of the attention economy, Apple is now working to make \textit{privacy} a selling point \cite{Hall2019-jo}, e.g., by preventing developers from tracking users across multiple apps without their active consent \cite{Apple_Inc_undated-tg}. At the regulatory level, a recent review of dark patterns by Naraynan et al. notes that if the design community does not self-regulate by setting standards for itself, it may be regulated by more onerous standards set by others \cite{Narayanan2020-lk}. The U.S. Senate is currently considering how to regulate social media, with one bill that would make it illegal to $``$\textit{manipulate a user interface with the purpose or substantial effect of obscuring, subverting, or impairing user autonomy$"$} \cite{McKay2019-nd} and another that would ban autoplay and infinite scroll \cite{Chen_undated-mj}. For designers, the language of dark patterns is an important way to contribute to a broader critical discussion of design practices in the the technology industry \cite{Gray2018-zz}. \par

We caution that the message of attention capture dark patterns should \textit{not} be $``$never X,$"$  but rather $``$be careful when X.$"$ Participants in both of our studies reported mixed experiences with many design mechanisms, including autoplay and recommendations. An outright ban on these mechanisms is likely to reduce sense of agency in a substantial number of situations where the user just wants to explore. Instead, a nuanced guide to dark patterns might present examples of the problem, followed by counterexamples where such a pattern is appropriate. While this creates a murky gray middle, it also better describes the effects of the design mechanisms that we identified in our studies.\par

\subsection{Limitations}
In addition to the previously stated limitations of our participant sampling and focus on design mechanisms as a unit of analysis, our work also has at least four conceptual limitations that could be explored in future work. First, both of our studies asked participants to share their preferences, however present bias \cite{ODonoghue2015-ue} predicts that \textit{actual behavior} will favor the choice that offers immediate rewards at the expense of long-term goals. An \textit{in-situ} study of how people respond to redesigns intended to influence sense of agency would yield results on (``what users do''), which might need to be reconciled with the present results on (``what users say''). Second, time and attention are not the only factors that influence sense of agency. By asking participants in both studies to reflect on\textit{ $``$...in control of how you spend your time on YouTube$"$} we discouraged participants from considering other factors such as privacy \cite{Sundar2010-oh}. In Study 2, this may have primed participants to focus on sense of agency over other factors when evaluating which version of the mockup they preferred. Third, self-reported agency can be quite different from the facts of agency \cite{Coyle2012-qr,Moore2016-mq}. For example, many people continue to press `placebo buttons’ like the `close door button’ in their apartment’s elevator, even when doing so has no effect \cite{Paumgarten2014-nh}. There is therefore a concern that designs to increase sense of agency may be disconnected from actual ability to influence the world. Fourth, users are not the only stakeholders on YouTube, and it would be a mistake to optimize for their sense of agency alone. Google, creators, advertisers, and even society itself all have a stake in what happens on YouTube. For instance, radicalizing political videos can make viewers feel as if they have uncovered powerful conspiracies that were previously hidden from them \cite{Roose2019-rl}; to support sense of agency in this use case would be dangerous. User sense of agency needs to be integrated into larger design frameworks as one important consideration among many for social media apps. \par

\section{Conclusion}
Whereas a common approach to digital wellbeing is designing to reduce screentime, this work takes an alternative approach of designing to increase sense of agency. In two studies, we identify mechanisms within the YouTube mobile app that participants report influence their sense of agency and how they want to change them. We find that participants generally prefer mechanisms like autoplay and recommendations to be redesigned for a greater sense of agency than the YouTube mobile app currently provides. For digital wellbeing designers, we highlight a need for recommender systems that better reflect user aspirations rather than just reinforce their current behavior. We also propose mechanisms that support `microplanning,' making lightweight plans to guide a single session of use, to increase user sense of agency. Finally, we propose language that the design community might adopt to recognize design patterns that impose attentional harms upon the user.

\begin{acks}
This work was funded in part by National Science Foundation award \#1849955. We thank Xuecong Xu, Ming Yao Zheng, Kevin Kuo, Tejus Krishnan, Laura Meng, Linda Lai, and Stefania Druga for helping to conceptualize this study and design the mockups.
\end{acks}

\nocite{*}

\bibliographystyle{ACM-Reference-Format}
\bibliography{references}


\begin{thebibliography}{122}


\ifx \showCODEN    \undefined \def \showCODEN     #1{\unskip}     \fi
\ifx \showDOI      \undefined \def \showDOI       #1{#1}\fi
\ifx \showISBNx    \undefined \def \showISBNx     #1{\unskip}     \fi
\ifx \showISBNxiii \undefined \def \showISBNxiii  #1{\unskip}     \fi
\ifx \showISSN     \undefined \def \showISSN      #1{\unskip}     \fi
\ifx \showLCCN     \undefined \def \showLCCN      #1{\unskip}     \fi
\ifx \shownote     \undefined \def \shownote      #1{#1}          \fi
\ifx \showarticletitle \undefined \def \showarticletitle #1{#1}   \fi
\ifx \showURL      \undefined \def \showURL       {\relax}        \fi
\providecommand\bibfield[2]{#2}
\providecommand\bibinfo[2]{#2}
\providecommand\natexlab[1]{#1}
\providecommand\showeprint[2][]{arXiv:#2}

\bibitem[\protect\citeauthoryear{??}{noa}{[n.d.]}]%
        {noauthor_undated-zy}
 \bibinfo{year}{[n.d.]}\natexlab{}.
\newblock \bibinfo{title}{Take Control}.
\newblock
  \bibinfo{howpublished}{\url{https://www.humanetech.com/take-control}}.
\newblock
\urldef\tempurl%
\url{https://www.humanetech.com/take-control}
\showURL{%
\tempurl}
\newblock
\shownote{Accessed: 2020-8-3.}


\bibitem[\protect\citeauthoryear{Aagaard}{Aagaard}{2015}]%
        {Aagaard2015-rf}
\bibfield{author}{\bibinfo{person}{Jesper Aagaard}.}
  \bibinfo{year}{2015}\natexlab{}.
\newblock \showarticletitle{Drawn to distraction: A qualitative study of
  off-task use of educational technology}.
\newblock \bibinfo{journal}{\emph{Computers \& education}}
  \bibinfo{volume}{87} (\bibinfo{date}{Sept.} \bibinfo{year}{2015}),
  \bibinfo{pages}{90--97}.
\newblock
\showISSN{0360-1315}
\urldef\tempurl%
\url{https://doi.org/10.1016/j.compedu.2015.03.010}
\showDOI{\tempurl}


\bibitem[\protect\citeauthoryear{Agapie}{Agapie}{2020}]%
        {Agapie2020-ne}
\bibfield{author}{\bibinfo{person}{Elena Agapie}.}
  \bibinfo{year}{2020}\natexlab{}.
\newblock \emph{\bibinfo{title}{Designing for Human Supported {Evidence-Based}
  Planning}}.
\newblock \bibinfo{thesistype}{Ph.D. Dissertation}.
\newblock
\urldef\tempurl%
\url{https://digital.lib.washington.edu/researchworks/handle/1773/45709}
\showURL{%
\tempurl}


\bibitem[\protect\citeauthoryear{Ames}{Ames}{2013}]%
        {Ames2013-ja}
\bibfield{author}{\bibinfo{person}{Morgan~G Ames}.}
  \bibinfo{year}{2013}\natexlab{}.
\newblock \showarticletitle{Managing Mobile Multitasking: The Culture of
  iPhones on Stanford Campus}. In \bibinfo{booktitle}{\emph{Proceedings of the
  2013 Conference on Computer Supported Cooperative Work}} (San Antonio, Texas,
  USA) \emph{(\bibinfo{series}{CSCW '13})}. \bibinfo{publisher}{ACM},
  \bibinfo{address}{New York, NY, USA}, \bibinfo{pages}{1487--1498}.
\newblock
\showISBNx{9781450313315}
\urldef\tempurl%
\url{https://doi.org/10.1145/2441776.2441945}
\showDOI{\tempurl}


\bibitem[\protect\citeauthoryear{{Apple Inc}}{{Apple Inc}}{[n.d.]}]%
        {Apple_Inc_undated-tg}
\bibfield{author}{\bibinfo{person}{{Apple Inc}}.}
  \bibinfo{year}{[n.d.]}\natexlab{}.
\newblock \bibinfo{title}{Details for app privacy questions now available -
  News - Apple Developer}.
\newblock
  \bibinfo{howpublished}{\url{https://developer.apple.com/news/?id=hx9s63c5}}.
\newblock
\urldef\tempurl%
\url{https://developer.apple.com/news/?id=hx9s63c5}
\showURL{%
\tempurl}
\newblock
\shownote{Accessed: 2020-9-13.}


\bibitem[\protect\citeauthoryear{Baab-Muguira}{Baab-Muguira}{2017}]%
        {Baab-Muguira2017-jr}
\bibfield{author}{\bibinfo{person}{Catherine Baab-Muguira}.}
  \bibinfo{year}{2017}\natexlab{}.
\newblock \bibinfo{title}{The Stupidly Simple Productivity Hack Hiding In
  Microsoft Word}.
\newblock
  \bibinfo{howpublished}{\url{https://www.fastcompany.com/3068825/the-stupidly-simple-productivity-hack-hiding-in-microsoft-word}}.
\newblock
\urldef\tempurl%
\url{https://www.fastcompany.com/3068825/the-stupidly-simple-productivity-hack-hiding-in-microsoft-word}
\showURL{%
\tempurl}
\newblock
\shownote{Accessed: 2020-9-11.}


\bibitem[\protect\citeauthoryear{Bannon, Bowers, Carstensen, Hughes, Kuutti,
  Pycock, Rodden, Schmidt, Shapiro, Sharrock, and {Others}}{Bannon
  et~al\mbox{.}}{1994}]%
        {Bannon1994-iy}
\bibfield{author}{\bibinfo{person}{Liam Bannon}, \bibinfo{person}{John Bowers},
  \bibinfo{person}{Peter Carstensen}, \bibinfo{person}{John~A Hughes},
  \bibinfo{person}{K Kuutti}, \bibinfo{person}{James Pycock},
  \bibinfo{person}{Tom Rodden}, \bibinfo{person}{Kjeld Schmidt},
  \bibinfo{person}{Dan Shapiro}, \bibinfo{person}{Wes Sharrock}, {and}
  \bibinfo{person}{{Others}}.} \bibinfo{year}{1994}\natexlab{}.
\newblock \bibinfo{booktitle}{\emph{Informing {CSCW} system requirements}}.
\newblock \bibinfo{publisher}{Lancaster University}.
\newblock
\urldef\tempurl%
\url{https://www.forskningsdatabasen.dk/en/catalog/2185760093}
\showURL{%
\tempurl}


\bibitem[\protect\citeauthoryear{Baumer, Adams, Khovanskaya, Liao, Smith,
  Schwanda~Sosik, and Williams}{Baumer et~al\mbox{.}}{2013}]%
        {Baumer2013-qp}
\bibfield{author}{\bibinfo{person}{Eric P~S Baumer}, \bibinfo{person}{Phil
  Adams}, \bibinfo{person}{Vera~D Khovanskaya}, \bibinfo{person}{Tony~C Liao},
  \bibinfo{person}{Madeline~E Smith}, \bibinfo{person}{Victoria
  Schwanda~Sosik}, {and} \bibinfo{person}{Kaiton Williams}.}
  \bibinfo{year}{2013}\natexlab{}.
\newblock \showarticletitle{Limiting, Leaving, and ({Re)Lapsing}: An
  Exploration of Facebook Non-use Practices and Experiences}. In
  \bibinfo{booktitle}{\emph{Proceedings of the {SIGCHI} Conference on Human
  Factors in Computing Systems}} (Paris, France) \emph{(\bibinfo{series}{CHI
  '13})}. \bibinfo{publisher}{ACM}, \bibinfo{address}{New York, NY, USA},
  \bibinfo{pages}{3257--3266}.
\newblock
\showISBNx{9781450318990}
\urldef\tempurl%
\url{https://doi.org/10.1145/2470654.2466446}
\showDOI{\tempurl}


\bibitem[\protect\citeauthoryear{Baumer and Silberman}{Baumer and
  Silberman}{2011}]%
        {Baumer2011-ho}
\bibfield{author}{\bibinfo{person}{Eric P~S Baumer} {and}
  \bibinfo{person}{M~Six Silberman}.} \bibinfo{year}{2011}\natexlab{}.
\newblock \showarticletitle{When the implication is not to design
  (technology)}. In \bibinfo{booktitle}{\emph{Proceedings of the {SIGCHI}
  Conference on Human Factors in Computing Systems}}. \bibinfo{publisher}{ACM},
  \bibinfo{pages}{2271--2274}.
\newblock
\showISBNx{9781450302289}
\urldef\tempurl%
\url{https://doi.org/10.1145/1978942.1979275}
\showDOI{\tempurl}


\bibitem[\protect\citeauthoryear{Baumer, Sun, and Schaedler}{Baumer
  et~al\mbox{.}}{2018}]%
        {Baumer2018-gv}
\bibfield{author}{\bibinfo{person}{Eric P~S Baumer}, \bibinfo{person}{Rui Sun},
  {and} \bibinfo{person}{Peter Schaedler}.} \bibinfo{year}{2018}\natexlab{}.
\newblock \showarticletitle{Departing and Returning: Sense of Agency As an
  Organizing Concept for Understanding Social Media {Non/Use} Transitions}.
\newblock \bibinfo{journal}{\emph{Proc. ACM Hum. -Comput. Interact.}}
  \bibinfo{volume}{2}, \bibinfo{number}{CSCW} (\bibinfo{date}{Nov.}
  \bibinfo{year}{2018}), \bibinfo{pages}{23:1--23:19}.
\newblock
\showISSN{2573-0142}
\urldef\tempurl%
\url{https://doi.org/10.1145/3274292}
\showDOI{\tempurl}


\bibitem[\protect\citeauthoryear{Berberian, Sarrazin, Le~Blaye, and
  Haggard}{Berberian et~al\mbox{.}}{2012}]%
        {Berberian2012-gt}
\bibfield{author}{\bibinfo{person}{Bruno Berberian},
  \bibinfo{person}{Jean-Christophe Sarrazin}, \bibinfo{person}{Patrick
  Le~Blaye}, {and} \bibinfo{person}{Patrick Haggard}.}
  \bibinfo{year}{2012}\natexlab{}.
\newblock \showarticletitle{Automation technology and sense of control: a
  window on human agency}.
\newblock \bibinfo{journal}{\emph{PloS one}} \bibinfo{volume}{7},
  \bibinfo{number}{3} (\bibinfo{date}{March} \bibinfo{year}{2012}),
  \bibinfo{pages}{e34075}.
\newblock
\showISSN{1932-6203}
\urldef\tempurl%
\url{https://doi.org/10.1371/journal.pone.0034075}
\showDOI{\tempurl}


\bibitem[\protect\citeauthoryear{Boyatzis}{Boyatzis}{1998}]%
        {Boyatzis1998-wf}
\bibfield{author}{\bibinfo{person}{Richard~E Boyatzis}.}
  \bibinfo{year}{1998}\natexlab{}.
\newblock \bibinfo{booktitle}{\emph{Transforming Qualitative Information:
  Thematic Analysis and Code Development}}.
\newblock \bibinfo{publisher}{SAGE}.
\newblock
\showISBNx{9780761909613}
\urldef\tempurl%
\url{https://play.google.com/store/books/details?id=_rfClWRhIKAC}
\showURL{%
\tempurl}


\bibitem[\protect\citeauthoryear{Braun and Clarke}{Braun and Clarke}{2006}]%
        {Braun2006-ma}
\bibfield{author}{\bibinfo{person}{Virginia Braun} {and}
  \bibinfo{person}{Victoria Clarke}.} \bibinfo{year}{2006}\natexlab{}.
\newblock \showarticletitle{Using thematic analysis in psychology}.
\newblock \bibinfo{journal}{\emph{Qualitative research in psychology}}
  \bibinfo{volume}{3}, \bibinfo{number}{2} (\bibinfo{date}{Jan.}
  \bibinfo{year}{2006}), \bibinfo{pages}{77--101}.
\newblock
\showISSN{1478-0887}
\urldef\tempurl%
\url{https://doi.org/10.1191/1478088706qp063oa}
\showDOI{\tempurl}


\bibitem[\protect\citeauthoryear{Braun and Clarke}{Braun and Clarke}{2019}]%
        {Braun2019-af}
\bibfield{author}{\bibinfo{person}{Virginia Braun} {and}
  \bibinfo{person}{Victoria Clarke}.} \bibinfo{year}{2019}\natexlab{}.
\newblock \showarticletitle{Reflecting on reflexive thematic analysis}.
\newblock \bibinfo{journal}{\emph{Qualitative Research in Sport, Exercise and
  Health}} \bibinfo{volume}{11}, \bibinfo{number}{4} (\bibinfo{date}{Aug.}
  \bibinfo{year}{2019}), \bibinfo{pages}{589--597}.
\newblock
\showISSN{2159-676X}
\urldef\tempurl%
\url{https://doi.org/10.1080/2159676X.2019.1628806}
\showDOI{\tempurl}


\bibitem[\protect\citeauthoryear{Braun, Clarke, Hayfield, and Terry}{Braun
  et~al\mbox{.}}{2018}]%
        {Braun2018-th}
\bibfield{author}{\bibinfo{person}{Virginia Braun}, \bibinfo{person}{Victoria
  Clarke}, \bibinfo{person}{Nikki Hayfield}, {and} \bibinfo{person}{Gareth
  Terry}.} \bibinfo{year}{2018}\natexlab{}.
\newblock \showarticletitle{Thematic Analysis}.
\newblock In \bibinfo{booktitle}{\emph{Handbook of Research Methods in Health
  Social Sciences}}, \bibfield{editor}{\bibinfo{person}{Pranee Liamputtong}}
  (Ed.). \bibinfo{publisher}{Springer Singapore}, \bibinfo{address}{Singapore},
  \bibinfo{pages}{1--18}.
\newblock
\showISBNx{9789811027796}
\urldef\tempurl%
\url{https://doi.org/10.1007/978-981-10-2779-6\_103-1}
\showDOI{\tempurl}


\bibitem[\protect\citeauthoryear{Brignull and Darlington}{Brignull and
  Darlington}{[n.d.]}]%
        {Brignull_undated-pv}
\bibfield{author}{\bibinfo{person}{Harry Brignull} {and}
  \bibinfo{person}{Alexander Darlington}.} \bibinfo{year}{[n.d.]}\natexlab{}.
\newblock \bibinfo{title}{What are dark patterns?}
\newblock \bibinfo{howpublished}{\url{https://www.darkpatterns.org/}}.
\newblock
\urldef\tempurl%
\url{https://www.darkpatterns.org/}
\showURL{%
\tempurl}
\newblock
\shownote{Accessed: 2019-9-28.}


\bibitem[\protect\citeauthoryear{Bryan, Karlan, and Nelson}{Bryan
  et~al\mbox{.}}{2010}]%
        {Bryan2010-wt}
\bibfield{author}{\bibinfo{person}{Gharad Bryan}, \bibinfo{person}{Dean
  Karlan}, {and} \bibinfo{person}{Scott Nelson}.}
  \bibinfo{year}{2010}\natexlab{}.
\newblock \showarticletitle{Commitment Devices}.
\newblock \bibinfo{journal}{\emph{Annual review of economics}}
  \bibinfo{volume}{2}, \bibinfo{number}{1} (\bibinfo{date}{Sept.}
  \bibinfo{year}{2010}), \bibinfo{pages}{671--698}.
\newblock
\showISSN{1941-1383}
\urldef\tempurl%
\url{https://doi.org/10.1146/annurev.economics.102308.124324}
\showDOI{\tempurl}


\bibitem[\protect\citeauthoryear{Burr, Cristianini, and Ladyman}{Burr
  et~al\mbox{.}}{2018}]%
        {Burr2018-ut}
\bibfield{author}{\bibinfo{person}{Christopher Burr}, \bibinfo{person}{Nello
  Cristianini}, {and} \bibinfo{person}{James Ladyman}.}
  \bibinfo{year}{2018}\natexlab{}.
\newblock \showarticletitle{An Analysis of the Interaction Between Intelligent
  Software Agents and Human Users}.
\newblock \bibinfo{journal}{\emph{Minds and Machines}} \bibinfo{volume}{28},
  \bibinfo{number}{4} (\bibinfo{date}{Sept.} \bibinfo{year}{2018}),
  \bibinfo{pages}{735--774}.
\newblock
\showISSN{0924-6495}
\urldef\tempurl%
\url{https://doi.org/10.1007/s11023-018-9479-0}
\showDOI{\tempurl}


\bibitem[\protect\citeauthoryear{{calkuta}}{{calkuta}}{[n.d.]}]%
        {Calkuta_undated-af}
\bibfield{author}{\bibinfo{person}{{calkuta}}.}
  \bibinfo{year}{[n.d.]}\natexlab{}.
\newblock \bibinfo{title}{{DF} Tube (Distraction Free for YouTube)}.
\newblock
  \bibinfo{howpublished}{\url{https://chrome.google.com/webstore/detail/df-tube-distraction-free/mjdepdfccjgcndkmemponafgioodelna?hl=en}}.
\newblock
\urldef\tempurl%
\url{https://chrome.google.com/webstore/detail/df-tube-distraction-free/mjdepdfccjgcndkmemponafgioodelna?hl=en}
\showURL{%
\tempurl}
\newblock
\shownote{Accessed: 2020-8-3.}


\bibitem[\protect\citeauthoryear{Caplan}{Caplan}{2010}]%
        {Caplan2010-ox}
\bibfield{author}{\bibinfo{person}{Scott~E Caplan}.}
  \bibinfo{year}{2010}\natexlab{}.
\newblock \showarticletitle{Theory and measurement of generalized problematic
  Internet use: A two-step approach}.
\newblock \bibinfo{journal}{\emph{Computers in human behavior}}
  \bibinfo{volume}{26}, \bibinfo{number}{5} (\bibinfo{date}{Sept.}
  \bibinfo{year}{2010}), \bibinfo{pages}{1089--1097}.
\newblock
\showISSN{0747-5632}
\urldef\tempurl%
\url{https://doi.org/10.1016/j.chb.2010.03.012}
\showDOI{\tempurl}


\bibitem[\protect\citeauthoryear{Cash, Rae, Steel, and Winkler}{Cash
  et~al\mbox{.}}{2012}]%
        {Cash2012-wj}
\bibfield{author}{\bibinfo{person}{Hilarie Cash}, \bibinfo{person}{Cosette~D
  Rae}, \bibinfo{person}{Ann~H Steel}, {and} \bibinfo{person}{Alexander
  Winkler}.} \bibinfo{year}{2012}\natexlab{}.
\newblock \showarticletitle{Internet Addiction: A Brief Summary of Research and
  Practice}.
\newblock \bibinfo{journal}{\emph{Current psychiatry reviews}}
  \bibinfo{volume}{8}, \bibinfo{number}{4} (\bibinfo{date}{Nov.}
  \bibinfo{year}{2012}), \bibinfo{pages}{292--298}.
\newblock
\showISSN{1573-4005}
\urldef\tempurl%
\url{https://doi.org/10.2174/157340012803520513}
\showDOI{\tempurl}


\bibitem[\protect\citeauthoryear{Cecchinato, Cox, and Bird}{Cecchinato
  et~al\mbox{.}}{2017}]%
        {Cecchinato2017-yl}
\bibfield{author}{\bibinfo{person}{Marta~E Cecchinato}, \bibinfo{person}{Anna~L
  Cox}, {and} \bibinfo{person}{Jon Bird}.} \bibinfo{year}{2017}\natexlab{}.
\newblock \showarticletitle{Always On(line)?: User Experience of Smartwatches
  and their Role within {Multi-Device} Ecologies}. In
  \bibinfo{booktitle}{\emph{Proceedings of the 2017 {CHI} Conference on Human
  Factors in Computing Systems}}. \bibinfo{publisher}{ACM},
  \bibinfo{pages}{3557--3568}.
\newblock
\showISBNx{9781450346559}
\urldef\tempurl%
\url{https://doi.org/10.1145/3025453.3025538}
\showDOI{\tempurl}


\bibitem[\protect\citeauthoryear{Chen}{Chen}{[n.d.]}]%
        {Chen_undated-mj}
\bibfield{author}{\bibinfo{person}{Angela Chen}.}
  \bibinfo{year}{[n.d.]}\natexlab{}.
\newblock \showarticletitle{A new bill would ban making social media too
  addictive}.
\newblock \bibinfo{journal}{\emph{MIT Technology Review}}
  (\bibinfo{year}{[n.\,d.]}).
\newblock
\showISSN{0040-1692}
\urldef\tempurl%
\url{https://www.technologyreview.com/2019/07/30/133976/josh-hawley-social-media-addictive-design-legislation-smart-act-bill/}
\showURL{%
\tempurl}


\bibitem[\protect\citeauthoryear{Cheng, Burke, and Davis}{Cheng
  et~al\mbox{.}}{2019}]%
        {Cheng2019-me}
\bibfield{author}{\bibinfo{person}{Justin Cheng}, \bibinfo{person}{Moira
  Burke}, {and} \bibinfo{person}{Elena~Goetz Davis}.}
  \bibinfo{year}{2019}\natexlab{}.
\newblock \showarticletitle{Understanding Perceptions of Problematic Facebook
  Use: When People Experience Negative Life Impact and a Lack of Control}. In
  \bibinfo{booktitle}{\emph{Proceedings of the 2019 {CHI} Conference on Human
  Factors in Computing Systems}}. \bibinfo{publisher}{ACM},
  \bibinfo{pages}{199}.
\newblock
\showISBNx{9781450359702}
\urldef\tempurl%
\url{https://doi.org/10.1145/3290605.3300429}
\showDOI{\tempurl}


\bibitem[\protect\citeauthoryear{Cheng, Lo, and Leskovec}{Cheng
  et~al\mbox{.}}{2017}]%
        {Cheng2017-ge}
\bibfield{author}{\bibinfo{person}{Justin Cheng}, \bibinfo{person}{Caroline
  Lo}, {and} \bibinfo{person}{Jure Leskovec}.} \bibinfo{year}{2017}\natexlab{}.
\newblock \showarticletitle{Predicting Intent Using Activity Logs: How Goal
  Specificity and Temporal Range Affect User Behavior}. In
  \bibinfo{booktitle}{\emph{Proceedings of the 26th International Conference on
  World Wide Web Companion}}. \bibinfo{publisher}{International World Wide Web
  Conferences Steering Committee}, \bibinfo{pages}{593--601}.
\newblock
\showISBNx{9781450349147}
\urldef\tempurl%
\url{https://doi.org/10.1145/3041021.3054198}
\showDOI{\tempurl}


\bibitem[\protect\citeauthoryear{Cohen}{Cohen}{2012}]%
        {Cohen2012-qk}
\bibfield{author}{\bibinfo{person}{Julie~E Cohen}.}
  \bibinfo{year}{2012}\natexlab{}.
\newblock \showarticletitle{What privacy is for}.
\newblock \bibinfo{journal}{\emph{Harvard law review}}  \bibinfo{volume}{126}
  (\bibinfo{year}{2012}), \bibinfo{pages}{1904}.
\newblock
\showISSN{0017-811X}
\urldef\tempurl%
\url{https://heinonline.org/hol-cgi-bin/get_pdf.cgi?handle=hein.journals/hlr126&section=88&casa_token=WL8sq3iJD9YAAAAA:ET2vmSL-cBa99C528GgMfSdFRCGzFo-aMK6PIepmiwFmC6VWTYJsbwQTSAXRYoi42VexRjoT}
\showURL{%
\tempurl}


\bibitem[\protect\citeauthoryear{Collins, Cox, Bird, and
  Cornish-Tresstail}{Collins et~al\mbox{.}}{2014}]%
        {Collins2014-bv}
\bibfield{author}{\bibinfo{person}{Emily I~M Collins}, \bibinfo{person}{Anna~L
  Cox}, \bibinfo{person}{Jon Bird}, {and} \bibinfo{person}{Cassie
  Cornish-Tresstail}.} \bibinfo{year}{2014}\natexlab{}.
\newblock \showarticletitle{Barriers to Engagement with a Personal Informatics
  Productivity Tool}. In \bibinfo{booktitle}{\emph{Proceedings of the 26th
  Australian {Computer-Human} Interaction Conference on Designing Futures: The
  Future of Design}} (Sydney, New South Wales, Australia)
  \emph{(\bibinfo{series}{OzCHI '14})}. \bibinfo{publisher}{ACM},
  \bibinfo{address}{New York, NY, USA}, \bibinfo{pages}{370--379}.
\newblock
\showISBNx{9781450306539}
\urldef\tempurl%
\url{https://doi.org/10.1145/2686612.2686668}
\showDOI{\tempurl}


\bibitem[\protect\citeauthoryear{Covington, Adams, and Sargin}{Covington
  et~al\mbox{.}}{2016}]%
        {Covington2016-nf}
\bibfield{author}{\bibinfo{person}{Paul Covington}, \bibinfo{person}{Jay
  Adams}, {and} \bibinfo{person}{Emre Sargin}.}
  \bibinfo{year}{2016}\natexlab{}.
\newblock \showarticletitle{Deep Neural Networks for {YouTube}
  Recommendations}. In \bibinfo{booktitle}{\emph{Proceedings of the 10th {ACM}
  Conference on Recommender Systems}} (Boston, Massachusetts, USA)
  \emph{(\bibinfo{series}{RecSys '16})}. \bibinfo{publisher}{Association for
  Computing Machinery}, \bibinfo{address}{New York, NY, USA},
  \bibinfo{pages}{191--198}.
\newblock
\showISBNx{9781450340359}
\urldef\tempurl%
\url{https://doi.org/10.1145/2959100.2959190}
\showDOI{\tempurl}


\bibitem[\protect\citeauthoryear{Cox, Gould, Cecchinato, Iacovides, and
  Renfree}{Cox et~al\mbox{.}}{2016}]%
        {Cox2016-ts}
\bibfield{author}{\bibinfo{person}{Anna~L Cox}, \bibinfo{person}{Sandy J~J
  Gould}, \bibinfo{person}{Marta~E Cecchinato}, \bibinfo{person}{Ioanna
  Iacovides}, {and} \bibinfo{person}{Ian Renfree}.}
  \bibinfo{year}{2016}\natexlab{}.
\newblock \showarticletitle{Design Frictions for Mindful Interactions: The Case
  for Microboundaries}. In \bibinfo{booktitle}{\emph{Proceedings of the 2016
  {CHI} Conference Extended Abstracts on Human Factors in Computing Systems}}
  (San Jose, California, USA) \emph{(\bibinfo{series}{CHI EA '16})}.
  \bibinfo{publisher}{ACM}, \bibinfo{address}{New York, NY, USA},
  \bibinfo{pages}{1389--1397}.
\newblock
\showISBNx{9781450340823}
\urldef\tempurl%
\url{https://doi.org/10.1145/2851581.2892410}
\showDOI{\tempurl}


\bibitem[\protect\citeauthoryear{Coyle, Moore, Kristensson, Fletcher, and
  Blackwell}{Coyle et~al\mbox{.}}{2012}]%
        {Coyle2012-qr}
\bibfield{author}{\bibinfo{person}{David Coyle}, \bibinfo{person}{James Moore},
  \bibinfo{person}{Per~Ola Kristensson}, \bibinfo{person}{Paul Fletcher}, {and}
  \bibinfo{person}{Alan Blackwell}.} \bibinfo{year}{2012}\natexlab{}.
\newblock \showarticletitle{{I} did that! Measuring users' experience of agency
  in their own actions}. In \bibinfo{booktitle}{\emph{Proceedings of the
  {SIGCHI} Conference on Human Factors in Computing Systems}} (Austin, Texas,
  USA) \emph{(\bibinfo{series}{CHI '12})}. \bibinfo{publisher}{Association for
  Computing Machinery}, \bibinfo{address}{New York, NY, USA},
  \bibinfo{pages}{2025--2034}.
\newblock
\showISBNx{9781450310154}
\urldef\tempurl%
\url{https://doi.org/10.1145/2207676.2208350}
\showDOI{\tempurl}


\bibitem[\protect\citeauthoryear{Davis, Dinhopl, and Hiniker}{Davis
  et~al\mbox{.}}{2019}]%
        {Davis2019-au}
\bibfield{author}{\bibinfo{person}{Katie Davis}, \bibinfo{person}{Anja
  Dinhopl}, {and} \bibinfo{person}{Alexis Hiniker}.}
  \bibinfo{year}{2019}\natexlab{}.
\newblock \showarticletitle{`` Everything's the Phone'': Understanding the
  Phone's Supercharged Role in {Parent-Teen} Relationships}. In
  \bibinfo{booktitle}{\emph{Proceedings of the 2019 {CHI} Conference on Human
  Factors in Computing Systems}}. \bibinfo{publisher}{dl.acm.org},
  \bibinfo{pages}{227}.
\newblock
\urldef\tempurl%
\url{https://dl.acm.org/citation.cfm?id=3300457}
\showURL{%
\tempurl}


\bibitem[\protect\citeauthoryear{Dearden and Finlay}{Dearden and
  Finlay}{2006}]%
        {Dearden2006-xm}
\bibfield{author}{\bibinfo{person}{Andy Dearden} {and} \bibinfo{person}{Janet
  Finlay}.} \bibinfo{year}{2006}\natexlab{}.
\newblock \showarticletitle{Pattern Languages in {HCI}: A Critical Review}.
\newblock \bibinfo{journal}{\emph{Human--Computer Interaction}}
  \bibinfo{volume}{21}, \bibinfo{number}{1} (\bibinfo{date}{March}
  \bibinfo{year}{2006}), \bibinfo{pages}{49--102}.
\newblock
\showISSN{0737-0024}
\urldef\tempurl%
\url{https://doi.org/10.1207/s15327051hci2101\_3}
\showDOI{\tempurl}


\bibitem[\protect\citeauthoryear{Delaney and Lades}{Delaney and Lades}{2017}]%
        {Delaney2017-ld}
\bibfield{author}{\bibinfo{person}{Liam Delaney} {and}
  \bibinfo{person}{Leonhard~K Lades}.} \bibinfo{year}{2017}\natexlab{}.
\newblock \showarticletitle{Present bias and everyday self-control failures: a
  day reconstruction study}.
\newblock \bibinfo{journal}{\emph{Journal of behavioral decision making}}
  \bibinfo{volume}{30}, \bibinfo{number}{5} (\bibinfo{year}{2017}),
  \bibinfo{pages}{1157--1167}.
\newblock
\showISSN{0894-3257}
\urldef\tempurl%
\url{https://onlinelibrary.wiley.com/doi/abs/10.1002/bdm.2031?casa_token=8axLfIY-YlEAAAAA:Z65pChy95G1cvF2v600EYc6oqnxIGFaC5DQJteKnCuK5AQ4Nqkj_YnMbnrJB9KxqhtTJHN0NUtiLOOsI}
\showURL{%
\tempurl}


\bibitem[\protect\citeauthoryear{{Digital Wellness Warriors}}{{Digital Wellness
  Warriors}}{2018}]%
        {Digital_Wellness_Warriors2018-qh}
\bibfield{author}{\bibinfo{person}{{Digital Wellness Warriors}}.}
  \bibinfo{year}{2018}\natexlab{}.
\newblock \bibinfo{title}{Apple: let developers help iPhone users with mental
  wellbeing}.
\newblock
  \bibinfo{howpublished}{\url{https://www.change.org/p/apple-allow-digital-wellness-developers-to-help-ios-users}}.
\newblock
\urldef\tempurl%
\url{https://www.change.org/p/apple-allow-digital-wellness-developers-to-help-ios-users}
\showURL{%
\tempurl}
\newblock
\shownote{Accessed: 2020-8-27.}


\bibitem[\protect\citeauthoryear{Dixon}{Dixon}{2019}]%
        {Dixon2019-jd}
\bibfield{author}{\bibinfo{person}{Colin Dixon}.}
  \bibinfo{year}{2019}\natexlab{}.
\newblock \bibinfo{title}{Why shutter {YouTube} Leanback when there are many
  potential users?}
\newblock
  \bibinfo{howpublished}{\url{https://nscreenmedia.com/why-shutter-youtube-leanback-browser-experience-now/}}.
\newblock
\urldef\tempurl%
\url{https://nscreenmedia.com/why-shutter-youtube-leanback-browser-experience-now/}
\showURL{%
\tempurl}
\newblock
\shownote{Accessed: 2020-9-7.}


\bibitem[\protect\citeauthoryear{Duckworth, White, Matteucci, Shearer, and
  Gross}{Duckworth et~al\mbox{.}}{2016}]%
        {Duckworth2016-mq}
\bibfield{author}{\bibinfo{person}{Angela~L Duckworth},
  \bibinfo{person}{Rachel~E White}, \bibinfo{person}{Alyssa~J Matteucci},
  \bibinfo{person}{Annie Shearer}, {and} \bibinfo{person}{James~J Gross}.}
  \bibinfo{year}{2016}\natexlab{}.
\newblock \showarticletitle{A Stitch in Time: Strategic {Self-Control} in High
  School and College Students}.
\newblock \bibinfo{journal}{\emph{Journal of educational psychology}}
  \bibinfo{volume}{108}, \bibinfo{number}{3} (\bibinfo{date}{April}
  \bibinfo{year}{2016}), \bibinfo{pages}{329--341}.
\newblock
\showISSN{0022-0663}
\urldef\tempurl%
\url{https://doi.org/10.1037/edu0000062}
\showDOI{\tempurl}


\bibitem[\protect\citeauthoryear{Ekstrand and Willemsen}{Ekstrand and
  Willemsen}{2016}]%
        {Ekstrand2016-qy}
\bibfield{author}{\bibinfo{person}{Michael~D Ekstrand} {and}
  \bibinfo{person}{Martijn~C Willemsen}.} \bibinfo{year}{2016}\natexlab{}.
\newblock \showarticletitle{Behaviorism is Not Enough: Better Recommendations
  Through Listening to Users}. In \bibinfo{booktitle}{\emph{Proceedings of the
  10th {ACM} Conference on Recommender Systems}} (Boston, Massachusetts, USA)
  \emph{(\bibinfo{series}{RecSys '16})}. \bibinfo{publisher}{ACM},
  \bibinfo{address}{New York, NY, USA}, \bibinfo{pages}{221--224}.
\newblock
\showISBNx{9781450340359}
\urldef\tempurl%
\url{https://doi.org/10.1145/2959100.2959179}
\showDOI{\tempurl}


\bibitem[\protect\citeauthoryear{Felner, Adan, Price, Cowen, Lorion, and
  Ramos-McKay}{Felner et~al\mbox{.}}{1988}]%
        {Felner1988-yr}
\bibfield{author}{\bibinfo{person}{Robert Felner}, \bibinfo{person}{Angela
  Adan}, \bibinfo{person}{Richard Price}, \bibinfo{person}{E~L Cowen},
  \bibinfo{person}{R~P Lorion}, {and} \bibinfo{person}{J Ramos-McKay}.}
  \bibinfo{year}{1988}\natexlab{}.
\newblock \showarticletitle{14 Ounces of prevention: A casebook for
  practitioners}.
\newblock  (\bibinfo{year}{1988}).
\newblock


\bibitem[\protect\citeauthoryear{Fogg}{Fogg}{2009}]%
        {Fogg2009-sy}
\bibfield{author}{\bibinfo{person}{B~J Fogg}.} \bibinfo{year}{2009}\natexlab{}.
\newblock \showarticletitle{Creating persuasive technologies: an eight-step
  design process}. In \bibinfo{booktitle}{\emph{Proceedings of the 4th
  International Conference on Persuasive Technology}} (Claremont, California,
  USA) \emph{(\bibinfo{series}{Persuasive '09}, \bibinfo{number}{Article 44})}.
  \bibinfo{publisher}{Association for Computing Machinery},
  \bibinfo{address}{New York, NY, USA}, \bibinfo{pages}{1--6}.
\newblock
\showISBNx{9781605583761}
\urldef\tempurl%
\url{https://doi.org/10.1145/1541948.1542005}
\showDOI{\tempurl}


\bibitem[\protect\citeauthoryear{Forsyth}{Forsyth}{2008}]%
        {forsyth2008self}
\bibfield{author}{\bibinfo{person}{Donelson~R Forsyth}.}
  \bibinfo{year}{2008}\natexlab{}.
\newblock \showarticletitle{Self-serving bias}.
\newblock  (\bibinfo{year}{2008}).
\newblock


\bibitem[\protect\citeauthoryear{Gollwitzer and Sheeran}{Gollwitzer and
  Sheeran}{2006}]%
        {Gollwitzer2006-ua}
\bibfield{author}{\bibinfo{person}{Peter~M Gollwitzer} {and}
  \bibinfo{person}{Paschal Sheeran}.} \bibinfo{year}{2006}\natexlab{}.
\newblock \showarticletitle{Implementation Intentions and Goal Achievement: A
  Meta‐analysis of Effects and Processes}.
\newblock In \bibinfo{booktitle}{\emph{Advances in Experimental Social
  Psychology}}. Vol.~\bibinfo{volume}{38}. \bibinfo{publisher}{Academic Press},
  \bibinfo{pages}{69--119}.
\newblock
\urldef\tempurl%
\url{https://doi.org/10.1016/S0065-2601(06)38002-1}
\showDOI{\tempurl}


\bibitem[\protect\citeauthoryear{{Google}}{{Google}}{[n.d.]}]%
        {Google_undated-yh}
\bibfield{author}{\bibinfo{person}{{Google}}.}
  \bibinfo{year}{[n.d.]}\natexlab{}.
\newblock \bibinfo{title}{{YouTube} Leanback offers effortless viewing}.
\newblock
  \bibinfo{howpublished}{\url{https://youtube.googleblog.com/2010/07/youtube-leanback-offers-effortless.html}}.
\newblock
\urldef\tempurl%
\url{https://youtube.googleblog.com/2010/07/youtube-leanback-offers-effortless.html}
\showURL{%
\tempurl}
\newblock
\shownote{Accessed: 2020-9-12.}


\bibitem[\protect\citeauthoryear{Gray, Kou, Battles, Hoggatt, and Toombs}{Gray
  et~al\mbox{.}}{2018}]%
        {Gray2018-zz}
\bibfield{author}{\bibinfo{person}{Colin~M Gray}, \bibinfo{person}{Yubo Kou},
  \bibinfo{person}{Bryan Battles}, \bibinfo{person}{Joseph Hoggatt}, {and}
  \bibinfo{person}{Austin~L Toombs}.} \bibinfo{year}{2018}\natexlab{}.
\newblock \showarticletitle{The Dark (Patterns) Side of {UX} Design}. In
  \bibinfo{booktitle}{\emph{Proceedings of the 2018 {CHI} Conference on Human
  Factors in Computing Systems}} (Montreal QC, Canada)
  \emph{(\bibinfo{series}{CHI '18})}. \bibinfo{publisher}{ACM},
  \bibinfo{address}{New York, NY, USA}, \bibinfo{pages}{534:1--534:14}.
\newblock
\showISBNx{9781450356206}
\urldef\tempurl%
\url{https://doi.org/10.1145/3173574.3174108}
\showDOI{\tempurl}


\bibitem[\protect\citeauthoryear{Hall}{Hall}{2019}]%
        {Hall2019-jo}
\bibfield{author}{\bibinfo{person}{Zac Hall}.} \bibinfo{year}{2019}\natexlab{}.
\newblock \bibinfo{title}{Apple makes privacy extremely relatable in fun new
  iPhone ad - 9to5Mac}.
\newblock
  \bibinfo{howpublished}{\url{https://9to5mac.com/2019/03/14/iphone-privacy-ad/}}.
\newblock
\urldef\tempurl%
\url{https://9to5mac.com/2019/03/14/iphone-privacy-ad/}
\showURL{%
\tempurl}
\newblock
\shownote{Accessed: 2020-9-13.}


\bibitem[\protect\citeauthoryear{Harambam, Bountouridis, Makhortykh, and van
  Hoboken}{Harambam et~al\mbox{.}}{2019}]%
        {Harambam2019-qu}
\bibfield{author}{\bibinfo{person}{Jaron Harambam}, \bibinfo{person}{Dimitrios
  Bountouridis}, \bibinfo{person}{Mykola Makhortykh}, {and}
  \bibinfo{person}{Joris van Hoboken}.} \bibinfo{year}{2019}\natexlab{}.
\newblock \showarticletitle{Designing for the Better by Taking Users into
  Account: A Qualitative Evaluation of User Control Mechanisms in (News)
  Recommender Systems}. In \bibinfo{booktitle}{\emph{Proceedings of the 13th
  {ACM} Conference on Recommender Systems}} (Copenhagen, Denmark)
  \emph{(\bibinfo{series}{RecSys '19})}. \bibinfo{publisher}{ACM},
  \bibinfo{address}{New York, NY, USA}, \bibinfo{pages}{69--77}.
\newblock
\showISBNx{9781450362436}
\urldef\tempurl%
\url{https://doi.org/10.1145/3298689.3347014}
\showDOI{\tempurl}


\bibitem[\protect\citeauthoryear{Harmon and Mazmanian}{Harmon and
  Mazmanian}{2013}]%
        {Harmon2013-ja}
\bibfield{author}{\bibinfo{person}{Ellie Harmon} {and} \bibinfo{person}{Melissa
  Mazmanian}.} \bibinfo{year}{2013}\natexlab{}.
\newblock \showarticletitle{Stories of the Smartphone in Everyday Discourse:
  Conflict, Tension \& Instability}. In \bibinfo{booktitle}{\emph{Proceedings
  of the {SIGCHI} Conference on Human Factors in Computing Systems}} (Paris,
  France) \emph{(\bibinfo{series}{CHI '13})}. \bibinfo{publisher}{ACM},
  \bibinfo{address}{New York, NY, USA}, \bibinfo{pages}{1051--1060}.
\newblock
\showISBNx{9781450318990}
\urldef\tempurl%
\url{https://doi.org/10.1145/2470654.2466134}
\showDOI{\tempurl}


\bibitem[\protect\citeauthoryear{Hill, Widdicks, and Hazas}{Hill
  et~al\mbox{.}}{2020}]%
        {Hill2020-hn}
\bibfield{author}{\bibinfo{person}{Joshua Hill}, \bibinfo{person}{Kelly
  Widdicks}, {and} \bibinfo{person}{Mike Hazas}.}
  \bibinfo{year}{2020}\natexlab{}.
\newblock \showarticletitle{Mapping the Scope of Software Interventions for
  Moderate Internet Use on Mobile Devices}. In
  \bibinfo{booktitle}{\emph{Proceedings of the 7th International Conference on
  {ICT} for Sustainability}} (Bristol, United Kingdom)
  \emph{(\bibinfo{series}{ICT4S2020})}. \bibinfo{publisher}{Association for
  Computing Machinery}, \bibinfo{address}{New York, NY, USA},
  \bibinfo{pages}{204--212}.
\newblock
\showISBNx{9781450375955}
\urldef\tempurl%
\url{https://doi.org/10.1145/3401335.3401361}
\showDOI{\tempurl}


\bibitem[\protect\citeauthoryear{Hiniker, Heung, Hong, and Kientz}{Hiniker
  et~al\mbox{.}}{2018}]%
        {Hiniker2018-ra}
\bibfield{author}{\bibinfo{person}{Alexis Hiniker}, \bibinfo{person}{Sharon~S
  Heung}, \bibinfo{person}{Sungsoo~(ray) Hong}, {and} \bibinfo{person}{Julie~A
  Kientz}.} \bibinfo{year}{2018}\natexlab{}.
\newblock \showarticletitle{Coco's Videos: An Empirical Investigation of
  {Video-Player} Design Features and Children's Media Use}. In
  \bibinfo{booktitle}{\emph{Proceedings of the 2018 {CHI} Conference on Human
  Factors in Computing Systems}} (Montreal QC, Canada)
  \emph{(\bibinfo{series}{CHI '18}, \bibinfo{number}{Paper 254})}.
  \bibinfo{publisher}{Association for Computing Machinery},
  \bibinfo{address}{New York, NY, USA}, \bibinfo{pages}{1--13}.
\newblock
\showISBNx{9781450356206}
\urldef\tempurl%
\url{https://doi.org/10.1145/3173574.3173828}
\showDOI{\tempurl}


\bibitem[\protect\citeauthoryear{Hiniker, Hong, Kohno, and Kientz}{Hiniker
  et~al\mbox{.}}{2016a}]%
        {Hiniker2016-cs}
\bibfield{author}{\bibinfo{person}{Alexis Hiniker},
  \bibinfo{person}{Sungsoo~(ray) Hong}, \bibinfo{person}{Tadayoshi Kohno},
  {and} \bibinfo{person}{Julie~A Kientz}.} \bibinfo{year}{2016}\natexlab{a}.
\newblock \showarticletitle{{MyTime}: Designing and Evaluating an Intervention
  for Smartphone {Non-Use}}. In \bibinfo{booktitle}{\emph{Proceedings of the
  2016 {CHI} Conference on Human Factors in Computing Systems}}.
  \bibinfo{publisher}{ACM}, \bibinfo{pages}{4746--4757}.
\newblock
\showISBNx{9781450333627}
\urldef\tempurl%
\url{https://doi.org/10.1145/2858036.2858403}
\showDOI{\tempurl}


\bibitem[\protect\citeauthoryear{Hiniker, Lee, Sobel, and Choe}{Hiniker
  et~al\mbox{.}}{2017}]%
        {Hiniker2017-sx}
\bibfield{author}{\bibinfo{person}{Alexis Hiniker}, \bibinfo{person}{Bongshin
  Lee}, \bibinfo{person}{Kiley Sobel}, {and} \bibinfo{person}{Eun~Kyoung
  Choe}.} \bibinfo{year}{2017}\natexlab{}.
\newblock \showarticletitle{Plan \& Play: Supporting Intentional Media Use in
  Early Childhood}. In \bibinfo{booktitle}{\emph{Proceedings of the 2017
  Conference on Interaction Design and Children}} (Stanford, California, USA)
  \emph{(\bibinfo{series}{IDC '17})}. \bibinfo{publisher}{ACM},
  \bibinfo{address}{New York, NY, USA}, \bibinfo{pages}{85--95}.
\newblock
\showISBNx{9781450349215}
\urldef\tempurl%
\url{https://doi.org/10.1145/3078072.3079752}
\showDOI{\tempurl}


\bibitem[\protect\citeauthoryear{Hiniker, Patel, Kohno, and Kientz}{Hiniker
  et~al\mbox{.}}{2016b}]%
        {Hiniker2016-ef}
\bibfield{author}{\bibinfo{person}{Alexis Hiniker}, \bibinfo{person}{Shwetak~N
  Patel}, \bibinfo{person}{Tadayoshi Kohno}, {and} \bibinfo{person}{Julie~A
  Kientz}.} \bibinfo{year}{2016}\natexlab{b}.
\newblock \showarticletitle{Why would you do that? predicting the uses and
  gratifications behind smartphone-usage behaviors}. In
  \bibinfo{booktitle}{\emph{Proceedings of the 2016 {ACM} International Joint
  Conference on Pervasive and Ubiquitous Computing}}.
  \bibinfo{publisher}{dl.acm.org}, \bibinfo{pages}{634--645}.
\newblock
\urldef\tempurl%
\url{http://dl.acm.org/citation.cfm?id=2971762}
\showURL{%
\tempurl}


\bibitem[\protect\citeauthoryear{Hofmann, Baumeister, F{\"o}rster, and
  Vohs}{Hofmann et~al\mbox{.}}{2012}]%
        {Hofmann2012-uv}
\bibfield{author}{\bibinfo{person}{Wilhelm Hofmann}, \bibinfo{person}{Roy~F
  Baumeister}, \bibinfo{person}{Georg F{\"o}rster}, {and}
  \bibinfo{person}{Kathleen~D Vohs}.} \bibinfo{year}{2012}\natexlab{}.
\newblock \showarticletitle{Everyday temptations: an experience sampling study
  of desire, conflict, and self-control}.
\newblock \bibinfo{journal}{\emph{Journal of personality and social
  psychology}} \bibinfo{volume}{102}, \bibinfo{number}{6} (\bibinfo{date}{June}
  \bibinfo{year}{2012}), \bibinfo{pages}{1318--1335}.
\newblock
\showISSN{0022-3514, 1939-1315}
\urldef\tempurl%
\url{https://doi.org/10.1037/a0026545}
\showDOI{\tempurl}


\bibitem[\protect\citeauthoryear{J{\"a}{\"a}skel{\"a}inen}{J{\"a}{\"a}skel{\"a}inen}{2010}]%
        {Jaaskelainen2010-ts}
\bibfield{author}{\bibinfo{person}{Riitta J{\"a}{\"a}skel{\"a}inen}.}
  \bibinfo{year}{2010}\natexlab{}.
\newblock \showarticletitle{Think-aloud protocol}.
\newblock \bibinfo{journal}{\emph{Handbook of translation studies}}
  \bibinfo{volume}{1} (\bibinfo{year}{2010}), \bibinfo{pages}{371--374}.
\newblock
\urldef\tempurl%
\url{https://books.google.com/books?hl=en&lr=&id=sBVGAYCh_9AC&oi=fnd&pg=PA371&dq=Riitta+J%C3%A4%C3%A4skel%C3%A4inen+2010+Think-aloud+protocol.&ots=Qn8NYacfXD&sig=9gAnKpZ6vdVbGHapKTfo74b0MtE}
\showURL{%
\tempurl}


\bibitem[\protect\citeauthoryear{Jeong, Kim, Yum, and Hwang}{Jeong
  et~al\mbox{.}}{2016}]%
        {Jeong2016-cg}
\bibfield{author}{\bibinfo{person}{Se-Hoon Jeong}, \bibinfo{person}{Hyoungjee
  Kim}, \bibinfo{person}{Jung-Yoon Yum}, {and} \bibinfo{person}{Yoori Hwang}.}
  \bibinfo{year}{2016}\natexlab{}.
\newblock \showarticletitle{What type of content are smartphone users addicted
  to?: {SNS} vs. games}.
\newblock \bibinfo{journal}{\emph{Computers in human behavior}}
  \bibinfo{volume}{54} (\bibinfo{date}{Jan.} \bibinfo{year}{2016}),
  \bibinfo{pages}{10--17}.
\newblock
\showISSN{0747-5632}
\urldef\tempurl%
\url{https://doi.org/10.1016/j.chb.2015.07.035}
\showDOI{\tempurl}


\bibitem[\protect\citeauthoryear{Kamenetz}{Kamenetz}{2018}]%
        {Kamenetz2018-sl}
\bibfield{author}{\bibinfo{person}{Anya Kamenetz}.}
  \bibinfo{year}{2018}\natexlab{}.
\newblock \bibinfo{booktitle}{\emph{The art of screen time: How your family can
  balance digital media and real life}}.
\newblock \bibinfo{publisher}{Hachette UK}.
\newblock


\bibitem[\protect\citeauthoryear{Kim, Jung, Ko, and Lee}{Kim
  et~al\mbox{.}}{2019a}]%
        {Kim2019-pq}
\bibfield{author}{\bibinfo{person}{Jaejeung Kim}, \bibinfo{person}{Hayoung
  Jung}, \bibinfo{person}{Minsam Ko}, {and} \bibinfo{person}{Uichin Lee}.}
  \bibinfo{year}{2019}\natexlab{a}.
\newblock \showarticletitle{{GoalKeeper}: Exploring Interaction Lockout
  Mechanisms for Regulating Smartphone Use}.
\newblock \bibinfo{journal}{\emph{Proc. ACM Interact. Mob. Wearable Ubiquitous
  Technol.}} \bibinfo{volume}{3}, \bibinfo{number}{1} (\bibinfo{date}{March}
  \bibinfo{year}{2019}), \bibinfo{pages}{29}.
\newblock
\urldef\tempurl%
\url{https://doi.org/10.1145/3314403}
\showDOI{\tempurl}


\bibitem[\protect\citeauthoryear{Kim, Park, Lee, Ko, and Lee}{Kim
  et~al\mbox{.}}{2019b}]%
        {Kim2019-cd}
\bibfield{author}{\bibinfo{person}{Jaejeung Kim}, \bibinfo{person}{Joonyoung
  Park}, \bibinfo{person}{Hyunsoo Lee}, \bibinfo{person}{Minsam Ko}, {and}
  \bibinfo{person}{Uichin Lee}.} \bibinfo{year}{2019}\natexlab{b}.
\newblock \showarticletitle{{LocknType}: Lockout Task Intervention for
  Discouraging Smartphone App Use,``}. In \bibinfo{booktitle}{\emph{{ACM}
  {CHI}}}.
\newblock
\urldef\tempurl%
\url{https://doi.org/10.1145/3290605.3300927}
\showDOI{\tempurl}


\bibitem[\protect\citeauthoryear{Kim, Jeon, Choe, Lee, Kim, and Seo}{Kim
  et~al\mbox{.}}{2016}]%
        {Kim2016-pj}
\bibfield{author}{\bibinfo{person}{Young-Ho Kim}, \bibinfo{person}{Jae~Ho
  Jeon}, \bibinfo{person}{Eun~Kyoung Choe}, \bibinfo{person}{Bongshin Lee},
  \bibinfo{person}{Kwonhyun Kim}, {and} \bibinfo{person}{Jinwook Seo}.}
  \bibinfo{year}{2016}\natexlab{}.
\newblock \showarticletitle{{TimeAware}: Leveraging Framing Effects to Enhance
  Personal Productivity}. In \bibinfo{booktitle}{\emph{Proceedings of the 2016
  {CHI} Conference on Human Factors in Computing Systems}} (San Jose,
  California, USA) \emph{(\bibinfo{series}{CHI '16})}.
  \bibinfo{publisher}{ACM}, \bibinfo{address}{New York, NY, USA},
  \bibinfo{pages}{272--283}.
\newblock
\showISBNx{9781450333627}
\urldef\tempurl%
\url{https://doi.org/10.1145/2858036.2858428}
\showDOI{\tempurl}


\bibitem[\protect\citeauthoryear{Ko, Yang, Lee, Heizmann, Jeong, Lee, Shin,
  Yatani, Song, and Chung}{Ko et~al\mbox{.}}{2015}]%
        {Ko2015-yc}
\bibfield{author}{\bibinfo{person}{Minsam Ko}, \bibinfo{person}{Subin Yang},
  \bibinfo{person}{Joonwon Lee}, \bibinfo{person}{Christian Heizmann},
  \bibinfo{person}{Jinyoung Jeong}, \bibinfo{person}{Uichin Lee},
  \bibinfo{person}{Daehee Shin}, \bibinfo{person}{Koji Yatani},
  \bibinfo{person}{Junehwa Song}, {and} \bibinfo{person}{Kyong-Mee Chung}.}
  \bibinfo{year}{2015}\natexlab{}.
\newblock \showarticletitle{{NUGU}: A Group-based Intervention App for
  Improving {Self-Regulation} of Limiting Smartphone Use}. In
  \bibinfo{booktitle}{\emph{Proceedings of the 18th {ACM} Conference on
  Computer Supported Cooperative Work \& Social Computing}} (Vancouver, BC,
  Canada) \emph{(\bibinfo{series}{CSCW '15})}. \bibinfo{publisher}{ACM},
  \bibinfo{address}{New York, NY, USA}, \bibinfo{pages}{1235--1245}.
\newblock
\showISBNx{9781450329224}
\urldef\tempurl%
\url{https://doi.org/10.1145/2675133.2675244}
\showDOI{\tempurl}


\bibitem[\protect\citeauthoryear{Kovacs, Gregory, Ma, Wu, Emami, Ray, and
  Bernstein}{Kovacs et~al\mbox{.}}{2019}]%
        {Kovacs2019-pe}
\bibfield{author}{\bibinfo{person}{Geza Kovacs}, \bibinfo{person}{Drew~Mylander
  Gregory}, \bibinfo{person}{Zilin Ma}, \bibinfo{person}{Zhengxuan Wu},
  \bibinfo{person}{Golrokh Emami}, \bibinfo{person}{Jacob Ray}, {and}
  \bibinfo{person}{Michael~S Bernstein}.} \bibinfo{year}{2019}\natexlab{}.
\newblock \showarticletitle{Conservation of Procrastination: Do Productivity
  Interventions Save Time Or Just Redistribute It?}. In
  \bibinfo{booktitle}{\emph{Proceedings of the 2019 {CHI} Conference on Human
  Factors in Computing Systems}} (Glasgow, Scotland Uk)
  \emph{(\bibinfo{series}{CHI '19})}. \bibinfo{publisher}{ACM},
  \bibinfo{address}{New York, NY, USA}, \bibinfo{pages}{330:1--330:12}.
\newblock
\showISBNx{9781450359702}
\urldef\tempurl%
\url{https://doi.org/10.1145/3290605.3300560}
\showDOI{\tempurl}


\bibitem[\protect\citeauthoryear{Landis and Koch}{Landis and Koch}{1977}]%
        {Landis1977-xl}
\bibfield{author}{\bibinfo{person}{J~R Landis} {and} \bibinfo{person}{G~G
  Koch}.} \bibinfo{year}{1977}\natexlab{}.
\newblock \showarticletitle{The measurement of observer agreement for
  categorical data}.
\newblock \bibinfo{journal}{\emph{Biometrics}} \bibinfo{volume}{33},
  \bibinfo{number}{1} (\bibinfo{date}{March} \bibinfo{year}{1977}),
  \bibinfo{pages}{159--174}.
\newblock
\showISSN{0006-341X}
\urldef\tempurl%
\url{https://www.ncbi.nlm.nih.gov/pubmed/843571}
\showURL{%
\tempurl}


\bibitem[\protect\citeauthoryear{Latham and Locke}{Latham and Locke}{1991}]%
        {Latham1991-nr}
\bibfield{author}{\bibinfo{person}{Gary~P Latham} {and}
  \bibinfo{person}{Edwin~A Locke}.} \bibinfo{year}{1991}\natexlab{}.
\newblock \showarticletitle{Self-regulation through goal setting}.
\newblock \bibinfo{journal}{\emph{Organizational behavior and human decision
  processes}} \bibinfo{volume}{50}, \bibinfo{number}{2} (\bibinfo{year}{1991}),
  \bibinfo{pages}{212--247}.
\newblock
\showISSN{0749-5978}
\urldef\tempurl%
\url{https://www.researchgate.net/profile/Gary_Latham2/publication/232501090_A_Theory_of_Goal_Setting_Task_Performance/links/57d0e85108ae5f03b489170d/A-Theory-of-Goal-Setting-Task-Performance.pdf}
\showURL{%
\tempurl}


\bibitem[\protect\citeauthoryear{Lewis}{Lewis}{2017}]%
        {Lewis2017-dr}
\bibfield{author}{\bibinfo{person}{Paul Lewis}.}
  \bibinfo{year}{2017}\natexlab{}.
\newblock \showarticletitle{'Our minds can be hijacked': the tech insiders who
  fear a smartphone dystopia}.
\newblock \bibinfo{journal}{\emph{The Guardian}} \bibinfo{volume}{6},
  \bibinfo{number}{10} (\bibinfo{year}{2017}), \bibinfo{pages}{2017}.
\newblock
\showISSN{0261-3077}


\bibitem[\protect\citeauthoryear{Limerick, Coyle, and Moore}{Limerick
  et~al\mbox{.}}{2014}]%
        {Limerick2014-lk}
\bibfield{author}{\bibinfo{person}{Hannah Limerick}, \bibinfo{person}{David
  Coyle}, {and} \bibinfo{person}{James~W Moore}.}
  \bibinfo{year}{2014}\natexlab{}.
\newblock \showarticletitle{The experience of agency in human-computer
  interactions: a review}.
\newblock \bibinfo{journal}{\emph{Frontiers in human neuroscience}}
  \bibinfo{volume}{8} (\bibinfo{date}{Aug.} \bibinfo{year}{2014}),
  \bibinfo{pages}{643}.
\newblock
\showISSN{1662-5161}
\urldef\tempurl%
\url{https://doi.org/10.3389/fnhum.2014.00643}
\showDOI{\tempurl}


\bibitem[\protect\citeauthoryear{Limerick, Moore, and Coyle}{Limerick
  et~al\mbox{.}}{2015}]%
        {Limerick2015-ui}
\bibfield{author}{\bibinfo{person}{Hannah Limerick}, \bibinfo{person}{James~W
  Moore}, {and} \bibinfo{person}{David Coyle}.}
  \bibinfo{year}{2015}\natexlab{}.
\newblock \showarticletitle{Empirical Evidence for a Diminished Sense of Agency
  in Speech Interfaces}. In \bibinfo{booktitle}{\emph{Proceedings of the 33rd
  Annual {ACM} Conference on Human Factors in Computing Systems}} (Seoul,
  Republic of Korea) \emph{(\bibinfo{series}{CHI '15})}.
  \bibinfo{publisher}{Association for Computing Machinery},
  \bibinfo{address}{New York, NY, USA}, \bibinfo{pages}{3967--3970}.
\newblock
\showISBNx{9781450331456}
\urldef\tempurl%
\url{https://doi.org/10.1145/2702123.2702379}
\showDOI{\tempurl}


\bibitem[\protect\citeauthoryear{Lottridge, Marschner, Wang, Romanovsky, and
  Nass}{Lottridge et~al\mbox{.}}{2012}]%
        {Lottridge2012-tj}
\bibfield{author}{\bibinfo{person}{Danielle Lottridge}, \bibinfo{person}{Eli
  Marschner}, \bibinfo{person}{Ellen Wang}, \bibinfo{person}{Maria Romanovsky},
  {and} \bibinfo{person}{Clifford Nass}.} \bibinfo{year}{2012}\natexlab{}.
\newblock \showarticletitle{Browser Design Impacts Multitasking}.
\newblock \bibinfo{journal}{\emph{Proceedings of the Human Factors and
  Ergonomics Society ... Annual Meeting Human Factors and Ergonomics Society.
  Meeting}} \bibinfo{volume}{56}, \bibinfo{number}{1} (\bibinfo{date}{Sept.}
  \bibinfo{year}{2012}), \bibinfo{pages}{1957--1961}.
\newblock
\showISSN{1541-9312}
\urldef\tempurl%
\url{https://doi.org/10.1177/1071181312561289}
\showDOI{\tempurl}


\bibitem[\protect\citeauthoryear{Lukoff, Hiniker, Gray, Mathur, and
  Chivukula}{Lukoff et~al\mbox{.}}{2021}]%
        {Lukoff2021-ru}
\bibfield{author}{\bibinfo{person}{Kai Lukoff}, \bibinfo{person}{Alexis
  Hiniker}, \bibinfo{person}{Colin~M Gray}, \bibinfo{person}{Arunesh Mathur},
  {and} \bibinfo{person}{Shruthi Chivukula}.} \bibinfo{year}{2021}\natexlab{}.
\newblock \showarticletitle{What Can {CHI} Do About Dark Patterns?}. In
  \bibinfo{booktitle}{\emph{Extended Abstracts of the 2021 {CHI} Conference on
  Human Factors in Computing Systems}} (Yokohama, Japan).
  \bibinfo{publisher}{ACM}, \bibinfo{address}{New York, NY, USA}.
\newblock
\urldef\tempurl%
\url{https://doi.org/10.1145/3411763.3441360}
\showDOI{\tempurl}


\bibitem[\protect\citeauthoryear{Lukoff, Yu, Kientz, and Hiniker}{Lukoff
  et~al\mbox{.}}{2018}]%
        {Lukoff2018-km}
\bibfield{author}{\bibinfo{person}{Kai Lukoff}, \bibinfo{person}{Cissy Yu},
  \bibinfo{person}{Julie Kientz}, {and} \bibinfo{person}{Alexis Hiniker}.}
  \bibinfo{year}{2018}\natexlab{}.
\newblock \showarticletitle{What Makes Smartphone Use Meaningful or
  Meaningless?}
\newblock \bibinfo{journal}{\emph{Proc. ACM Interact. Mob. Wearable Ubiquitous
  Technol.}} \bibinfo{volume}{2}, \bibinfo{number}{1} (\bibinfo{date}{March}
  \bibinfo{year}{2018}), \bibinfo{pages}{22:1--22:26}.
\newblock
\showISSN{2474-9567}
\urldef\tempurl%
\url{https://doi.org/10.1145/3191754}
\showDOI{\tempurl}


\bibitem[\protect\citeauthoryear{Lyngs, Binns, Van~Kleek, and {others}}{Lyngs
  et~al\mbox{.}}{2018}]%
        {Lyngs2018-qi}
\bibfield{author}{\bibinfo{person}{U Lyngs}, \bibinfo{person}{R Binns},
  \bibinfo{person}{M Van~Kleek}, {and} \bibinfo{person}{{others}}.}
  \bibinfo{year}{2018}\natexlab{}.
\newblock \showarticletitle{So, Tell Me What Users Want, What They Really,
  Really Want!}
\newblock \bibinfo{journal}{\emph{Extended Abstracts of the}}
  (\bibinfo{year}{2018}).
\newblock
\urldef\tempurl%
\url{https://dl.acm.org/citation.cfm?id=3188397}
\showURL{%
\tempurl}


\bibitem[\protect\citeauthoryear{Lyngs, Lukoff, Slovak, Binns, Slack, Inzlicht,
  Van~Leek, and Shadbolt}{Lyngs et~al\mbox{.}}{2019}]%
        {Lyngs2019-mw}
\bibfield{author}{\bibinfo{person}{Ulrik Lyngs}, \bibinfo{person}{Kai Lukoff},
  \bibinfo{person}{Petr Slovak}, \bibinfo{person}{Reuben Binns},
  \bibinfo{person}{Adam Slack}, \bibinfo{person}{Michael Inzlicht},
  \bibinfo{person}{Max Van~Leek}, {and} \bibinfo{person}{Nigel Shadbolt}.}
  \bibinfo{year}{2019}\natexlab{}.
\newblock \showarticletitle{{Self-Control} in Cyberspace: Applying Dual Systems
  Theory to a Review of Digital {Self-Control} Tools}.
\newblock \bibinfo{journal}{\emph{CHI 2019}} (\bibinfo{date}{May}
  \bibinfo{year}{2019}).
\newblock
\urldef\tempurl%
\url{https://doi.org/10.1145/3290605.3300361}
\showDOI{\tempurl}


\bibitem[\protect\citeauthoryear{Lyngs, Lukoff, Slovak, Seymour, Webb, Jirotka,
  Van~Kleek, and Shadbolt}{Lyngs et~al\mbox{.}}{2020}]%
        {Lyngs2020-pm}
\bibfield{author}{\bibinfo{person}{Ulrik Lyngs}, \bibinfo{person}{Kai Lukoff},
  \bibinfo{person}{Petr Slovak}, \bibinfo{person}{William Seymour},
  \bibinfo{person}{Helena Webb}, \bibinfo{person}{Marina Jirotka},
  \bibinfo{person}{Max Van~Kleek}, {and} \bibinfo{person}{Nigel Shadbolt}.}
  \bibinfo{year}{2020}\natexlab{}.
\newblock \showarticletitle{'I Just Want to Hack Myself to Not Get Distracted':
  Evaluating Design Interventions for {Self-Control} on Facebook}.
\newblock  (\bibinfo{date}{Jan.} \bibinfo{year}{2020}).
\newblock
\showeprint[arxiv]{2001.04180}~[cs.HC]
\urldef\tempurl%
\url{http://arxiv.org/abs/2001.04180}
\showURL{%
\tempurl}


\bibitem[\protect\citeauthoryear{Marino, Gini, Vieno, and Spada}{Marino
  et~al\mbox{.}}{2018}]%
        {Marino2018-qb}
\bibfield{author}{\bibinfo{person}{Claudia Marino}, \bibinfo{person}{Gianluca
  Gini}, \bibinfo{person}{Alessio Vieno}, {and} \bibinfo{person}{Marcantonio~M
  Spada}.} \bibinfo{year}{2018}\natexlab{}.
\newblock \showarticletitle{A comprehensive meta-analysis on Problematic
  Facebook Use}.
\newblock \bibinfo{journal}{\emph{Computers in human behavior}}
  \bibinfo{volume}{83} (\bibinfo{date}{June} \bibinfo{year}{2018}),
  \bibinfo{pages}{262--277}.
\newblock
\showISSN{0747-5632}
\urldef\tempurl%
\url{https://doi.org/10.1016/j.chb.2018.02.009}
\showDOI{\tempurl}


\bibitem[\protect\citeauthoryear{Marotta and Acquisti}{Marotta and
  Acquisti}{[n.d.]}]%
        {Marotta_undated-xi}
\bibfield{author}{\bibinfo{person}{Veronica Marotta} {and}
  \bibinfo{person}{Alessandro Acquisti}.} \bibinfo{year}{[n.d.]}\natexlab{}.
\newblock \showarticletitle{Online Distractions, Website Blockers, and Economic
  Productivity: A Randomized Field Experiment}.
\newblock  (\bibinfo{year}{[n.\,d.]}).
\newblock


\bibitem[\protect\citeauthoryear{Matney}{Matney}{2017}]%
        {Matney2017-ac}
\bibfield{author}{\bibinfo{person}{Lucas Matney}.}
  \bibinfo{year}{2017}\natexlab{}.
\newblock \showarticletitle{{YouTube} has 1.5 billion logged-in monthly users
  watching a ton of mobile video}.
\newblock \bibinfo{journal}{\emph{TechCrunch}} (\bibinfo{date}{June}
  \bibinfo{year}{2017}).
\newblock
\urldef\tempurl%
\url{http://techcrunch.com/2017/06/22/youtube-has-1-5-billion-logged-in-monthly-users-watching-a-ton-of-mobile-video/}
\showURL{%
\tempurl}


\bibitem[\protect\citeauthoryear{McKay}{McKay}{2019}]%
        {McKay2019-nd}
\bibfield{author}{\bibinfo{person}{Tom McKay}.}
  \bibinfo{year}{2019}\natexlab{}.
\newblock \bibinfo{title}{Senators Introduce Bill to Stop 'Dark Patterns' Huge
  Platforms Use to Trick Users}.
\newblock
  \bibinfo{howpublished}{\url{https://gizmodo.com/senators-introduce-bill-to-stop-dark-patterns-huge-plat-1833929276}}.
\newblock
\urldef\tempurl%
\url{https://gizmodo.com/senators-introduce-bill-to-stop-dark-patterns-huge-plat-1833929276}
\showURL{%
\tempurl}
\newblock
\shownote{Accessed: 2020-8-27.}


\bibitem[\protect\citeauthoryear{McNee, Lam, Guetzlaff, Konstan, and
  Riedl}{McNee et~al\mbox{.}}{2003}]%
        {McNee2003-ni}
\bibfield{author}{\bibinfo{person}{Sean~M McNee}, \bibinfo{person}{Shyong~K
  Lam}, \bibinfo{person}{Catherine Guetzlaff}, \bibinfo{person}{Joseph~A
  Konstan}, {and} \bibinfo{person}{John Riedl}.}
  \bibinfo{year}{2003}\natexlab{}.
\newblock \showarticletitle{Confidence displays and training in recommender
  systems}. In \bibinfo{booktitle}{\emph{Proc. {INTERACT}}},
  Vol.~\bibinfo{volume}{3}. \bibinfo{publisher}{books.google.com},
  \bibinfo{pages}{176--183}.
\newblock
\urldef\tempurl%
\url{https://books.google.com/books?hl=en&lr=&id=PTg0fVYqgCcC&oi=fnd&pg=PA176&dq=low+confidence+%22recommender+system%22&ots=ObJIxzmCAZ&sig=0uf4-fGfAJyBGbfmCDJE0zpPsGU}
\showURL{%
\tempurl}


\bibitem[\protect\citeauthoryear{McNee, Riedl, and Konstan}{McNee
  et~al\mbox{.}}{2006}]%
        {McNee2006-ji}
\bibfield{author}{\bibinfo{person}{Sean~M McNee}, \bibinfo{person}{John Riedl},
  {and} \bibinfo{person}{Joseph~A Konstan}.} \bibinfo{year}{2006}\natexlab{}.
\newblock \showarticletitle{Being accurate is not enough: how accuracy metrics
  have hurt recommender systems}. In \bibinfo{booktitle}{\emph{{CHI} '06
  Extended Abstracts on Human Factors in Computing Systems}} (Montr{\'e}al,
  Qu{\'e}bec, Canada) \emph{(\bibinfo{series}{CHI EA '06})}.
  \bibinfo{publisher}{Association for Computing Machinery},
  \bibinfo{address}{New York, NY, USA}, \bibinfo{pages}{1097--1101}.
\newblock
\showISBNx{9781595932983}
\urldef\tempurl%
\url{https://doi.org/10.1145/1125451.1125659}
\showDOI{\tempurl}


\bibitem[\protect\citeauthoryear{Metcalfe and Greene}{Metcalfe and
  Greene}{2007}]%
        {Metcalfe2007-cp}
\bibfield{author}{\bibinfo{person}{Janet Metcalfe} {and}
  \bibinfo{person}{Matthew~Jason Greene}.} \bibinfo{year}{2007}\natexlab{}.
\newblock \showarticletitle{Metacognition of agency}.
\newblock \bibinfo{journal}{\emph{Journal of experimental psychology. General}}
  \bibinfo{volume}{136}, \bibinfo{number}{2} (\bibinfo{date}{May}
  \bibinfo{year}{2007}), \bibinfo{pages}{184--199}.
\newblock
\showISSN{0096-3445, 0022-1015}
\urldef\tempurl%
\url{https://doi.org/10.1037/0096-3445.136.2.184}
\showDOI{\tempurl}


\bibitem[\protect\citeauthoryear{Monge~Roffarello and
  De~Russis}{Monge~Roffarello and De~Russis}{2019}]%
        {Monge_Roffarello2019-xt}
\bibfield{author}{\bibinfo{person}{Alberto Monge~Roffarello} {and}
  \bibinfo{person}{Luigi De~Russis}.} \bibinfo{year}{2019}\natexlab{}.
\newblock \showarticletitle{The Race Towards Digital Wellbeing: Issues and
  Opportunities}. In \bibinfo{booktitle}{\emph{Proceedings of the 2019 {CHI}
  Conference on Human Factors in Computing Systems}} (Glasgow, Scotland Uk)
  \emph{(\bibinfo{series}{CHI '19})}. \bibinfo{publisher}{ACM},
  \bibinfo{address}{New York, NY, USA}, \bibinfo{pages}{386:1--386:14}.
\newblock
\showISBNx{9781450359702}
\urldef\tempurl%
\url{https://doi.org/10.1145/3290605.3300616}
\showDOI{\tempurl}


\bibitem[\protect\citeauthoryear{Moore}{Moore}{2016}]%
        {Moore2016-mq}
\bibfield{author}{\bibinfo{person}{James~W Moore}.}
  \bibinfo{year}{2016}\natexlab{}.
\newblock \showarticletitle{What Is the Sense of Agency and Why Does it
  Matter?}
\newblock \bibinfo{journal}{\emph{Frontiers in psychology}}
  \bibinfo{volume}{7} (\bibinfo{date}{Aug.} \bibinfo{year}{2016}),
  \bibinfo{pages}{1272}.
\newblock
\showISSN{1664-1078}
\urldef\tempurl%
\url{https://doi.org/10.3389/fpsyg.2016.01272}
\showDOI{\tempurl}


\bibitem[\protect\citeauthoryear{Moser, Schoenebeck, and Resnick}{Moser
  et~al\mbox{.}}{2019}]%
        {Moser2019-vp}
\bibfield{author}{\bibinfo{person}{C Moser}, \bibinfo{person}{S~Y Schoenebeck},
  {and} \bibinfo{person}{P Resnick}.} \bibinfo{year}{2019}\natexlab{}.
\newblock \showarticletitle{Impulse Buying: Design Practices and Consumer
  Needs}.
\newblock \bibinfo{journal}{\emph{of the 2019 CHI Conference on …}}
  (\bibinfo{year}{2019}).
\newblock
\urldef\tempurl%
\url{https://dl.acm.org/doi/abs/10.1145/3290605.3300472}
\showURL{%
\tempurl}


\bibitem[\protect\citeauthoryear{Narayanan, Mathur, Chetty, and
  Kshirsagar}{Narayanan et~al\mbox{.}}{2020}]%
        {Narayanan2020-lk}
\bibfield{author}{\bibinfo{person}{Arvind Narayanan}, \bibinfo{person}{Arunesh
  Mathur}, \bibinfo{person}{Marshini Chetty}, {and} \bibinfo{person}{Mihir
  Kshirsagar}.} \bibinfo{year}{2020}\natexlab{}.
\newblock \showarticletitle{Dark Patterns: Past, Present, and Future: The
  evolution of tricky user interfaces}.
\newblock \bibinfo{journal}{\emph{Queueing Systems. Theory and Applications}}
  \bibinfo{volume}{18}, \bibinfo{number}{2} (\bibinfo{date}{April}
  \bibinfo{year}{2020}), \bibinfo{pages}{67--92}.
\newblock
\showISSN{0257-0130, 1542-7730}
\urldef\tempurl%
\url{https://doi.org/10.1145/3400899.3400901}
\showDOI{\tempurl}


\bibitem[\protect\citeauthoryear{Nielsen}{Nielsen}{1994}]%
        {Nielsen1994-ss}
\bibfield{author}{\bibinfo{person}{Jakob Nielsen}.}
  \bibinfo{year}{1994}\natexlab{}.
\newblock \bibinfo{title}{10 Heuristics for User Interface Design: Article by
  Jakob Nielsen}.
\newblock
  \bibinfo{howpublished}{\url{https://www.nngroup.com/articles/ten-usability-heuristics/}}.
\newblock
\urldef\tempurl%
\url{https://www.nngroup.com/articles/ten-usability-heuristics/}
\showURL{%
\tempurl}
\newblock
\shownote{Accessed: 2020-2-7.}


\bibitem[\protect\citeauthoryear{{NPR}}{{NPR}}{2015}]%
        {Npr2015-ps}
\bibfield{author}{\bibinfo{person}{{NPR}}.} \bibinfo{year}{2015}\natexlab{}.
\newblock \showarticletitle{Episode 653: The {Anti-Store}}.
\newblock \bibinfo{journal}{\emph{NPR}} (\bibinfo{date}{Sept.}
  \bibinfo{year}{2015}).
\newblock
\urldef\tempurl%
\url{https://www.npr.org/sections/money/2015/09/25/443519599/episode-653-the-anti-store}
\showURL{%
\tempurl}


\bibitem[\protect\citeauthoryear{O'Donoghue and Rabin}{O'Donoghue and
  Rabin}{2015}]%
        {ODonoghue2015-ue}
\bibfield{author}{\bibinfo{person}{Ted O'Donoghue} {and}
  \bibinfo{person}{Matthew Rabin}.} \bibinfo{year}{2015}\natexlab{}.
\newblock \showarticletitle{Present Bias: Lessons Learned and to Be Learned}.
\newblock \bibinfo{journal}{\emph{The American economic review}}
  \bibinfo{volume}{105}, \bibinfo{number}{5} (\bibinfo{year}{2015}),
  \bibinfo{pages}{273--279}.
\newblock
\showISSN{0002-8282}
\urldef\tempurl%
\url{https://ideas.repec.org/a/aea/aecrev/v105y2015i5p273-79.html}
\showURL{%
\tempurl}


\bibitem[\protect\citeauthoryear{Okeke, Sobolev, Dell, and Estrin}{Okeke
  et~al\mbox{.}}{2018}]%
        {Okeke2018-ad}
\bibfield{author}{\bibinfo{person}{Fabian Okeke}, \bibinfo{person}{Michael
  Sobolev}, \bibinfo{person}{Nicola Dell}, {and} \bibinfo{person}{Deborah
  Estrin}.} \bibinfo{year}{2018}\natexlab{}.
\newblock \showarticletitle{Good vibrations: can a digital nudge reduce digital
  overload?}. In \bibinfo{booktitle}{\emph{Proceedings of the 20th
  International Conference on {Human-Computer} Interaction with Mobile Devices
  and Services}}. \bibinfo{publisher}{ACM}, \bibinfo{pages}{4}.
\newblock
\showISBNx{9781450358989}
\urldef\tempurl%
\url{https://doi.org/10.1145/3229434.3229463}
\showDOI{\tempurl}


\bibitem[\protect\citeauthoryear{Oulasvirta, Rattenbury, Ma, and
  Raita}{Oulasvirta et~al\mbox{.}}{2012}]%
        {Oulasvirta2012-cc}
\bibfield{author}{\bibinfo{person}{Antti Oulasvirta}, \bibinfo{person}{Tye
  Rattenbury}, \bibinfo{person}{Lingyi Ma}, {and} \bibinfo{person}{Eeva
  Raita}.} \bibinfo{year}{2012}\natexlab{}.
\newblock \showarticletitle{Habits make smartphone use more pervasive}.
\newblock \bibinfo{journal}{\emph{Personal and Ubiquitous Computing}}
  \bibinfo{volume}{16}, \bibinfo{number}{1} (\bibinfo{date}{Jan.}
  \bibinfo{year}{2012}), \bibinfo{pages}{105--114}.
\newblock
\showISSN{0949-2054}
\urldef\tempurl%
\url{https://doi.org/10.1007/s00779-011-0412-2}
\showDOI{\tempurl}


\bibitem[\protect\citeauthoryear{Pandey}{Pandey}{2017}]%
        {Pandey2017-it}
\bibfield{author}{\bibinfo{person}{Erica Pandey}.}
  \bibinfo{year}{2017}\natexlab{}.
\newblock \bibinfo{title}{Sean Parker: Facebook was designed to exploit human
  ``vulnerability''}.
\newblock
  \bibinfo{howpublished}{\url{https://www.axios.com/sean-parker-facebook-exploits-a-vulnerability-in-humans-2507917325.html}}.
\newblock
\urldef\tempurl%
\url{https://www.axios.com/sean-parker-facebook-exploits-a-vulnerability-in-humans-2507917325.html}
\showURL{%
\tempurl}
\newblock
\shownote{Accessed: 2020-9-15.}


\bibitem[\protect\citeauthoryear{Park, Sim, Kim, Yi, and Lee}{Park
  et~al\mbox{.}}{2018}]%
        {Park2018-xs}
\bibfield{author}{\bibinfo{person}{Joonyoung Park}, \bibinfo{person}{Jin~Yong
  Sim}, \bibinfo{person}{Jaejeung Kim}, \bibinfo{person}{Mun~Yong Yi}, {and}
  \bibinfo{person}{Uichin Lee}.} \bibinfo{year}{2018}\natexlab{}.
\newblock \showarticletitle{Interaction Restraint: Enforcing Adaptive Cognitive
  Tasks to Restrain Problematic User Interaction}. In
  \bibinfo{booktitle}{\emph{Extended Abstracts of the 2018 {CHI} Conference on
  Human Factors in Computing Systems}} (Montreal QC, Canada)
  \emph{(\bibinfo{series}{CHI EA '18})}. \bibinfo{publisher}{ACM},
  \bibinfo{address}{New York, NY, USA}, \bibinfo{pages}{LBW559:1--LBW559:6}.
\newblock
\showISBNx{9781450356213}
\urldef\tempurl%
\url{https://doi.org/10.1145/3170427.3188613}
\showDOI{\tempurl}


\bibitem[\protect\citeauthoryear{Paumgarten}{Paumgarten}{2014}]%
        {Paumgarten2014-nh}
\bibfield{author}{\bibinfo{person}{Nick Paumgarten}.}
  \bibinfo{year}{2014}\natexlab{}.
\newblock \showarticletitle{Up And Then Down}.
\newblock \bibinfo{journal}{\emph{The New Yorker}} (\bibinfo{date}{July}
  \bibinfo{year}{2014}).
\newblock
\showISSN{0028-792X}
\urldef\tempurl%
\url{https://www.newyorker.com/magazine/2008/04/21/up-and-then-down}
\showURL{%
\tempurl}


\bibitem[\protect\citeauthoryear{Perrin and Anderson}{Perrin and
  Anderson}{2019}]%
        {Perrin2019-jr}
\bibfield{author}{\bibinfo{person}{Andrew Perrin} {and} \bibinfo{person}{Monica
  Anderson}.} \bibinfo{year}{2019}\natexlab{}.
\newblock \bibinfo{title}{Share of {U.S}. adults using social media, including
  Facebook, is mostly unchanged since 2018}.
\newblock
  \bibinfo{howpublished}{\url{https://www.pewresearch.org/fact-tank/2019/04/10/share-of-u-s-adults-using-social-media-including-facebook-is-mostly-unchanged-since-2018/}}.
\newblock
\urldef\tempurl%
\url{https://www.pewresearch.org/fact-tank/2019/04/10/share-of-u-s-adults-using-social-media-including-facebook-is-mostly-unchanged-since-2018/}
\showURL{%
\tempurl}
\newblock
\shownote{Accessed: 2020-9-14.}


\bibitem[\protect\citeauthoryear{Pinder, Vermeulen, Cowan, and {others}}{Pinder
  et~al\mbox{.}}{2018}]%
        {Pinder2018-uk}
\bibfield{author}{\bibinfo{person}{C Pinder}, \bibinfo{person}{J Vermeulen},
  \bibinfo{person}{B~R Cowan}, {and} \bibinfo{person}{{others}}.}
  \bibinfo{year}{2018}\natexlab{}.
\newblock \showarticletitle{Digital Behaviour Change Interventions to Break and
  Form Habits}.
\newblock \bibinfo{journal}{\emph{ACM Transactions on}} (\bibinfo{year}{2018}).
\newblock
\urldef\tempurl%
\url{https://dl.acm.org/citation.cfm?id=3196830}
\showURL{%
\tempurl}


\bibitem[\protect\citeauthoryear{Resnick and Varian}{Resnick and
  Varian}{1997}]%
        {Resnick1997-rd}
\bibfield{author}{\bibinfo{person}{Paul Resnick} {and} \bibinfo{person}{Hal~R
  Varian}.} \bibinfo{year}{1997}\natexlab{}.
\newblock \showarticletitle{Recommender systems}.
\newblock \bibinfo{journal}{\emph{Commun. ACM}} \bibinfo{volume}{40},
  \bibinfo{number}{3} (\bibinfo{year}{1997}), \bibinfo{pages}{56--58}.
\newblock
\showISSN{0001-0782}
\urldef\tempurl%
\url{http://dl.acm.org/citation.cfm?id=245121}
\showURL{%
\tempurl}


\bibitem[\protect\citeauthoryear{Roose}{Roose}{2019}]%
        {Roose2019-rl}
\bibfield{author}{\bibinfo{person}{Kevin Roose}.}
  \bibinfo{year}{2019}\natexlab{}.
\newblock \showarticletitle{The making of a {YouTube} radical}.
\newblock \bibinfo{journal}{\emph{The New York times}} (\bibinfo{year}{2019}).
\newblock
\showISSN{0362-4331}
\urldef\tempurl%
\url{https://static01.nyt.com/images/2019/06/09/nytfrontpage/scan.pdf}
\showURL{%
\tempurl}


\bibitem[\protect\citeauthoryear{Rubin}{Rubin}{1984}]%
        {Rubin1984-so}
\bibfield{author}{\bibinfo{person}{Alan~M Rubin}.}
  \bibinfo{year}{1984}\natexlab{}.
\newblock \showarticletitle{Ritualized and Instrumental Television Viewing}.
\newblock \bibinfo{journal}{\emph{The Journal of communication}}
  \bibinfo{volume}{34}, \bibinfo{number}{3} (\bibinfo{date}{Sept.}
  \bibinfo{year}{1984}), \bibinfo{pages}{67--77}.
\newblock
\showISSN{0021-9916, 1460-2466}
\urldef\tempurl%
\url{https://doi.org/10.1111/j.1460-2466.1984.tb02174.x}
\showDOI{\tempurl}


\bibitem[\protect\citeauthoryear{Ryan and Deci}{Ryan and Deci}{2006}]%
        {Ryan2006-lb}
\bibfield{author}{\bibinfo{person}{Richard~M Ryan} {and}
  \bibinfo{person}{Edward~L Deci}.} \bibinfo{year}{2006}\natexlab{}.
\newblock \showarticletitle{Self-regulation and the problem of human autonomy:
  Does psychology need choice, self-determination, and will?}
\newblock \bibinfo{journal}{\emph{Journal of personality}}
  \bibinfo{volume}{74}, \bibinfo{number}{6} (\bibinfo{year}{2006}),
  \bibinfo{pages}{1557--1586}.
\newblock
\showISSN{0022-3506}
\urldef\tempurl%
\url{https://onlinelibrary.wiley.com/doi/abs/10.1111/j.1467-6494.2006.00420.x?casa_token=MZuzC4Br_U4AAAAA:ErU7WbByAWUFcoh2N_5TIRqe7jhVXe6V8Z0-pWB8gbb-ZZ3I8xz_qrdAtePASmiTFBWb2COF6sX4BQ}
\showURL{%
\tempurl}


\bibitem[\protect\citeauthoryear{Schelling}{Schelling}{1984}]%
        {Schelling1984-zo}
\bibfield{author}{\bibinfo{person}{Thomas~C Schelling}.}
  \bibinfo{year}{1984}\natexlab{}.
\newblock \showarticletitle{{Self-Command} in Practice, in Policy, and in a
  Theory of Rational Choice}.
\newblock \bibinfo{journal}{\emph{The American economic review}}
  \bibinfo{volume}{74}, \bibinfo{number}{2} (\bibinfo{year}{1984}),
  \bibinfo{pages}{1--11}.
\newblock
\showISSN{0002-8282}
\urldef\tempurl%
\url{http://www.jstor.org/stable/1816322}
\showURL{%
\tempurl}


\bibitem[\protect\citeauthoryear{Schlosser}{Schlosser}{2019}]%
        {Schlosser2019-vl}
\bibfield{author}{\bibinfo{person}{Markus Schlosser}.}
  \bibinfo{year}{2019}\natexlab{}.
\newblock \bibinfo{booktitle}{\emph{Agency} (\bibinfo{edition}{winter 2019}
  ed.)}.
\newblock \bibinfo{publisher}{Metaphysics Research Lab, Stanford University}.
\newblock
\urldef\tempurl%
\url{https://plato.stanford.edu/archives/win2019/entries/agency/}
\showURL{%
\tempurl}


\bibitem[\protect\citeauthoryear{Schrage}{Schrage}{1996}]%
        {Schrage1996-hh}
\bibfield{author}{\bibinfo{person}{Michael Schrage}.}
  \bibinfo{year}{1996}\natexlab{}.
\newblock \showarticletitle{Cultures of prototyping}.
\newblock \bibinfo{journal}{\emph{Bringing design to software}}
  \bibinfo{volume}{4}, \bibinfo{number}{1} (\bibinfo{year}{1996}),
  \bibinfo{pages}{1--11}.
\newblock
\urldef\tempurl%
\url{https://pdfs.semanticscholar.org/c399/6e2738e52ea22c83ef662ab118f68b82eba2.pdf}
\showURL{%
\tempurl}


\bibitem[\protect\citeauthoryear{Sch{\"u}ll}{Sch{\"u}ll}{2012}]%
        {Schull2012-wx}
\bibfield{author}{\bibinfo{person}{N~D Sch{\"u}ll}.}
  \bibinfo{year}{2012}\natexlab{}.
\newblock \bibinfo{booktitle}{\emph{Addiction by Design: Machine Gambling in
  Las Vegas}}.
\newblock \bibinfo{publisher}{Princeton University Press}.
\newblock
\showISBNx{9780691127552}
\showLCCN{2012004339}
\urldef\tempurl%
\url{https://books.google.com/books?id=_Vsk6EXc1_4C}
\showURL{%
\tempurl}


\bibitem[\protect\citeauthoryear{Seaver}{Seaver}{2018}]%
        {Seaver2018-cn}
\bibfield{author}{\bibinfo{person}{Nick Seaver}.}
  \bibinfo{year}{2018}\natexlab{}.
\newblock \showarticletitle{Captivating algorithms: Recommender systems as
  traps}.
\newblock \bibinfo{journal}{\emph{Journal of Material Culture}}
  (\bibinfo{date}{Dec.} \bibinfo{year}{2018}),
  \bibinfo{pages}{1359183518820366}.
\newblock
\showISSN{1359-1835}
\urldef\tempurl%
\url{https://doi.org/10.1177/1359183518820366}
\showDOI{\tempurl}


\bibitem[\protect\citeauthoryear{Shneiderman}{Shneiderman}{1992}]%
        {Shneiderman1992-mo}
\bibfield{author}{\bibinfo{person}{Ben Shneiderman}.}
  \bibinfo{year}{1992}\natexlab{}.
\newblock \bibinfo{booktitle}{\emph{Designing the user interface (2nd ed.):
  strategies for effective human-computer interaction}}.
\newblock \bibinfo{publisher}{Addison-Wesley Longman Publishing Co., Inc.},
  \bibinfo{address}{USA}.
\newblock
\showISBNx{9780201572865}
\urldef\tempurl%
\url{https://dl.acm.org/citation.cfm?id=129385}
\showURL{%
\tempurl}


\bibitem[\protect\citeauthoryear{Shneiderman and Plaisant}{Shneiderman and
  Plaisant}{2004}]%
        {Shneiderman2004-yh}
\bibfield{author}{\bibinfo{person}{Ben Shneiderman} {and}
  \bibinfo{person}{Catherine Plaisant}.} \bibinfo{year}{2004}\natexlab{}.
\newblock \bibinfo{booktitle}{\emph{Designing the User Interface: Strategies
  for Effective {Human-Computer} Interaction (4th Edition)}}.
\newblock \bibinfo{publisher}{Pearson Addison Wesley}.
\newblock
\showISBNx{9780321197863}


\bibitem[\protect\citeauthoryear{Sillito}{Sillito}{[n.d.]}]%
        {Sillito_undated-xq}
\bibfield{author}{\bibinfo{person}{Jonathan Sillito}.}
  \bibinfo{year}{[n.d.]}\natexlab{}.
\newblock \bibinfo{title}{Saturate App: Simple Collaborative Analysis}.
\newblock \bibinfo{howpublished}{\url{http://www.saturateapp.com/}}.
\newblock
\urldef\tempurl%
\url{http://www.saturateapp.com/}
\showURL{%
\tempurl}
\newblock
\shownote{Accessed: 2020-2-NA.}


\bibitem[\protect\citeauthoryear{Silver, Smith, Johnson, Taylor, Jiang, Monica,
  and Rainie}{Silver et~al\mbox{.}}{2019}]%
        {Silver2019-xj}
\bibfield{author}{\bibinfo{person}{L Silver}, \bibinfo{person}{A Smith},
  \bibinfo{person}{C Johnson}, \bibinfo{person}{K Taylor}, \bibinfo{person}{J
  Jiang}, \bibinfo{person}{A Monica}, {and} \bibinfo{person}{L Rainie}.}
  \bibinfo{year}{2019}\natexlab{}.
\newblock \bibinfo{title}{Use of smartphones and social media is common across
  most emerging economies}.
\newblock
  \bibinfo{howpublished}{\url{https://www.pewresearch.org/internet/2019/03/07/use-of-smartphones-and-social-media-is-common-across-most-emerging-economies/}}.
\newblock
\urldef\tempurl%
\url{https://www.pewresearch.org/internet/2019/03/07/use-of-smartphones-and-social-media-is-common-across-most-emerging-economies/}
\showURL{%
\tempurl}
\newblock
\shownote{Accessed: 2019-2-NA.}


\bibitem[\protect\citeauthoryear{Smith, Toor, and Van~Kessel}{Smith
  et~al\mbox{.}}{2018}]%
        {Smith2018-av}
\bibfield{author}{\bibinfo{person}{Aaron Smith}, \bibinfo{person}{Skye Toor},
  {and} \bibinfo{person}{Patrick Van~Kessel}.} \bibinfo{year}{2018}\natexlab{}.
\newblock \bibinfo{title}{Many Turn to {YouTube} for Children's Content, News,
  {How-To} Lessons}.
\newblock
  \bibinfo{howpublished}{\url{https://www.pewresearch.org/internet/2018/11/07/many-turn-to-youtube-for-childrens-content-news-how-to-lessons/}}.
\newblock
\urldef\tempurl%
\url{https://www.pewresearch.org/internet/2018/11/07/many-turn-to-youtube-for-childrens-content-news-how-to-lessons/}
\showURL{%
\tempurl}
\newblock
\shownote{Accessed: 2020-3-3.}


\bibitem[\protect\citeauthoryear{Sniehotta, Scholz, and Schwarzer}{Sniehotta
  et~al\mbox{.}}{2005}]%
        {Sniehotta2005-wi}
\bibfield{author}{\bibinfo{person}{Falko~F Sniehotta}, \bibinfo{person}{Urte
  Scholz}, {and} \bibinfo{person}{Ralf Schwarzer}.}
  \bibinfo{year}{2005}\natexlab{}.
\newblock \showarticletitle{Bridging the intention--behaviour gap: Planning,
  self-efficacy, and action control in the adoption and maintenance of physical
  exercise}.
\newblock \bibinfo{journal}{\emph{Psychology \& Health}} \bibinfo{volume}{20},
  \bibinfo{number}{2} (\bibinfo{date}{April} \bibinfo{year}{2005}),
  \bibinfo{pages}{143--160}.
\newblock
\showISSN{0887-0446}
\urldef\tempurl%
\url{https://doi.org/10.1080/08870440512331317670}
\showDOI{\tempurl}


\bibitem[\protect\citeauthoryear{Solsman}{Solsman}{2018}]%
        {Solsman2018-wn}
\bibfield{author}{\bibinfo{person}{Joan~E Solsman}.}
  \bibinfo{year}{2018}\natexlab{}.
\newblock \bibinfo{title}{Ever get caught in an unexpected hourlong {YouTube}
  binge? Thank {YouTube} {AI} for that}.
\newblock
  \bibinfo{howpublished}{\url{https://www.cnet.com/news/youtube-ces-2018-neal-mohan/}}.
\newblock
\urldef\tempurl%
\url{https://www.cnet.com/news/youtube-ces-2018-neal-mohan/}
\showURL{%
\tempurl}
\newblock
\shownote{Accessed: 2020-5-1.}


\bibitem[\protect\citeauthoryear{Spangler}{Spangler}{[n.d.]}]%
        {Spangler_undated-my}
\bibfield{author}{\bibinfo{person}{Todd Spangler}.}
  \bibinfo{year}{[n.d.]}\natexlab{}.
\newblock \bibinfo{title}{{YouTube} Tops 20 Million Paying Subscribers,
  {YouTube} {TV} Has Over 2 Million Customers}.
\newblock
  \bibinfo{howpublished}{\url{https://variety.com/2020/digital/news/youtube-tops-20-million-paying-subscribers-youtube-tv-has-over-2-million-customers-1203491228/}}.
\newblock
\urldef\tempurl%
\url{https://variety.com/2020/digital/news/youtube-tops-20-million-paying-subscribers-youtube-tv-has-over-2-million-customers-1203491228/}
\showURL{%
\tempurl}
\newblock
\shownote{Accessed: 2020-8-26.}


\bibitem[\protect\citeauthoryear{Stanphill}{Stanphill}{2019}]%
        {Stanphill2019-gu}
\bibfield{author}{\bibinfo{person}{Maggie Stanphill}.}
  \bibinfo{year}{2019}\natexlab{}.
\newblock \bibinfo{title}{Optimizing for Engagement: Understanding the Use of
  Persuasive Technology on Internet Platforms}.
\newblock \bibinfo{howpublished}{U.S. Senate Committee Hearing}.
\newblock


\bibitem[\protect\citeauthoryear{Statt}{Statt}{2016}]%
        {Statt2016-uo}
\bibfield{author}{\bibinfo{person}{Nick Statt}.}
  \bibinfo{year}{2016}\natexlab{}.
\newblock \bibinfo{title}{Flowstate is a writing app that will delete
  everything if you stop typing}.
\newblock
  \bibinfo{howpublished}{\url{https://www.theverge.com/2016/1/28/10853534/flowstate-writing-app-mac-ios-delete-everything}}.
\newblock
\urldef\tempurl%
\url{https://www.theverge.com/2016/1/28/10853534/flowstate-writing-app-mac-ios-delete-everything}
\showURL{%
\tempurl}
\newblock
\shownote{Accessed: 2020-8-10.}


\bibitem[\protect\citeauthoryear{Sundar and Marathe}{Sundar and
  Marathe}{2010}]%
        {Sundar2010-oh}
\bibfield{author}{\bibinfo{person}{S~Shyam Sundar} {and}
  \bibinfo{person}{Sampada~S Marathe}.} \bibinfo{year}{2010}\natexlab{}.
\newblock \showarticletitle{Personalization versus Customization: the
  Importance of Agency, Privacy, and Power Usage}.
\newblock \bibinfo{journal}{\emph{Human communication research}}
  \bibinfo{volume}{36}, \bibinfo{number}{3} (\bibinfo{date}{July}
  \bibinfo{year}{2010}), \bibinfo{pages}{298--322}.
\newblock
\showISSN{0360-3989}
\urldef\tempurl%
\url{https://doi.org/10.1111/j.1468-2958.2010.01377.x}
\showDOI{\tempurl}


\bibitem[\protect\citeauthoryear{Synofzik, Vosgerau, and Newen}{Synofzik
  et~al\mbox{.}}{2008}]%
        {Synofzik2008-ia}
\bibfield{author}{\bibinfo{person}{Matthis Synofzik},
  \bibinfo{person}{Gottfried Vosgerau}, {and} \bibinfo{person}{Albert Newen}.}
  \bibinfo{year}{2008}\natexlab{}.
\newblock \showarticletitle{Beyond the comparator model: a multifactorial
  two-step account of agency}.
\newblock \bibinfo{journal}{\emph{Consciousness and cognition}}
  \bibinfo{volume}{17}, \bibinfo{number}{1} (\bibinfo{date}{March}
  \bibinfo{year}{2008}), \bibinfo{pages}{219--239}.
\newblock
\showISSN{1053-8100, 1090-2376}
\urldef\tempurl%
\url{https://doi.org/10.1016/j.concog.2007.03.010}
\showDOI{\tempurl}


\bibitem[\protect\citeauthoryear{Tapal, Oren, Dar, and Eitam}{Tapal
  et~al\mbox{.}}{2017}]%
        {Tapal2017-lk}
\bibfield{author}{\bibinfo{person}{Adam Tapal}, \bibinfo{person}{Ela Oren},
  \bibinfo{person}{Reuven Dar}, {and} \bibinfo{person}{Baruch Eitam}.}
  \bibinfo{year}{2017}\natexlab{}.
\newblock \showarticletitle{The Sense of Agency Scale: A Measure of Consciously
  Perceived Control over One's Mind, Body, and the Immediate Environment}.
\newblock \bibinfo{journal}{\emph{Frontiers in psychology}}
  \bibinfo{volume}{8} (\bibinfo{date}{Sept.} \bibinfo{year}{2017}),
  \bibinfo{pages}{1552}.
\newblock
\showISSN{1664-1078}
\urldef\tempurl%
\url{https://doi.org/10.3389/fpsyg.2017.01552}
\showDOI{\tempurl}


\bibitem[\protect\citeauthoryear{Tran, Yang, Davis, and Hiniker}{Tran
  et~al\mbox{.}}{2019}]%
        {Tran2019-lv}
\bibfield{author}{\bibinfo{person}{Jonathan~A Tran},
  \bibinfo{person}{Katherine~S Yang}, \bibinfo{person}{Katie Davis}, {and}
  \bibinfo{person}{Alexis Hiniker}.} \bibinfo{year}{2019}\natexlab{}.
\newblock \showarticletitle{Modeling the {Engagement-Disengagement} Cycle of
  Compulsive Phone Use}. In \bibinfo{booktitle}{\emph{{CHI} '19}} (Glasgow).
\newblock
\urldef\tempurl%
\url{https://doi.org/10.1145/3290605.3300542}
\showDOI{\tempurl}


\bibitem[\protect\citeauthoryear{{United States Census Bureau}}{{United States
  Census Bureau}}{[n.d.]}]%
        {United_States_Census_Bureau_undated-re}
\bibfield{author}{\bibinfo{person}{{United States Census Bureau}}.}
  \bibinfo{year}{[n.d.]}\natexlab{}.
\newblock \bibinfo{title}{{QuickFacts}: United States}.
\newblock
\newblock
\urldef\tempurl%
\url{https://www.census.gov/quickfacts/fact/table/US/PST045219}
\showURL{%
\tempurl}


\bibitem[\protect\citeauthoryear{Van~Kessel, Toor, and Smith}{Van~Kessel
  et~al\mbox{.}}{2019}]%
        {Van_Kessel2019-gn}
\bibfield{author}{\bibinfo{person}{Patrick Van~Kessel}, \bibinfo{person}{Skye
  Toor}, {and} \bibinfo{person}{Aaron Smith}.} \bibinfo{year}{2019}\natexlab{}.
\newblock \bibinfo{title}{A Week in the Life of Popular {YouTube} Channels}.
\newblock
  \bibinfo{howpublished}{\url{https://www.pewresearch.org/internet/2019/07/25/a-week-in-the-life-of-popular-youtube-channels/}}.
\newblock
\urldef\tempurl%
\url{https://www.pewresearch.org/internet/2019/07/25/a-week-in-the-life-of-popular-youtube-channels/}
\showURL{%
\tempurl}
\newblock
\shownote{Accessed: 2020-4-1.}


\bibitem[\protect\citeauthoryear{Williams}{Williams}{2018}]%
        {Williams2018-co}
\bibfield{author}{\bibinfo{person}{James Williams}.}
  \bibinfo{year}{2018}\natexlab{}.
\newblock \bibinfo{booktitle}{\emph{Stand Out of Our Light: Freedom and
  Resistance in the Attention Economy}}.
\newblock \bibinfo{publisher}{Cambridge University Press}.
\newblock
\showISBNx{9781108429092}
\urldef\tempurl%
\url{https://play.google.com/store/books/details?id=88FWDwAAQBAJ}
\showURL{%
\tempurl}


\bibitem[\protect\citeauthoryear{Wu, Pedersen, and Salehi}{Wu
  et~al\mbox{.}}{2019}]%
        {Wu2019-uc}
\bibfield{author}{\bibinfo{person}{Eva~Yiwei Wu}, \bibinfo{person}{Emily
  Pedersen}, {and} \bibinfo{person}{Niloufar Salehi}.}
  \bibinfo{year}{2019}\natexlab{}.
\newblock \showarticletitle{Agent, Gatekeeper, Drug Dealer: How Content
  Creators Craft Algorithmic Personas}.
\newblock \bibinfo{journal}{\emph{Proc. ACM Hum. -Comput. Interact.}}
  \bibinfo{volume}{3}, \bibinfo{number}{CSCW} (\bibinfo{date}{Nov.}
  \bibinfo{year}{2019}), \bibinfo{pages}{219:1--219:27}.
\newblock
\showISSN{2573-0142}
\urldef\tempurl%
\url{https://doi.org/10.1145/3359321}
\showDOI{\tempurl}


\bibitem[\protect\citeauthoryear{{YouTube}}{{YouTube}}{[n.d.]}]%
        {YouTube_undated-js}
\bibfield{author}{\bibinfo{person}{{YouTube}}.}
  \bibinfo{year}{[n.d.]}\natexlab{}.
\newblock \bibinfo{title}{{YouTube} for Press}.
\newblock \bibinfo{howpublished}{\url{https://www.youtube.com/about/press/}}.
\newblock
\urldef\tempurl%
\url{https://www.youtube.com/about/press/}
\showURL{%
\tempurl}
\newblock
\shownote{Accessed: 2020-8-14.}


\bibitem[\protect\citeauthoryear{Zagal, Bj{\"o}rk, and Lewis}{Zagal
  et~al\mbox{.}}{2013}]%
        {Zagal2013-nm}
\bibfield{author}{\bibinfo{person}{Jos{\'e}~P Zagal}, \bibinfo{person}{Staffan
  Bj{\"o}rk}, {and} \bibinfo{person}{Chris Lewis}.}
  \bibinfo{year}{2013}\natexlab{}.
\newblock \showarticletitle{Dark Patterns in the Design of Games}. In
  \bibinfo{booktitle}{\emph{Foundations of Digital Games 2013}}.
  \bibinfo{publisher}{diva-portal.org}.
\newblock
\urldef\tempurl%
\url{http://www.diva-portal.org/smash/record.jsf?pid=diva2:1043332}
\showURL{%
\tempurl}


\bibitem[\protect\citeauthoryear{Zimmerman and Forlizzi}{Zimmerman and
  Forlizzi}{2014}]%
        {Zimmerman2014-pu}
\bibfield{author}{\bibinfo{person}{John Zimmerman} {and} \bibinfo{person}{Jodi
  Forlizzi}.} \bibinfo{year}{2014}\natexlab{}.
\newblock \showarticletitle{Research Through Design in {HCI}}.
\newblock In \bibinfo{booktitle}{\emph{Ways of Knowing in {HCI}}},
  \bibfield{editor}{\bibinfo{person}{Judith~S Olson} {and}
  \bibinfo{person}{Wendy~A Kellogg}} (Eds.). \bibinfo{publisher}{Springer New
  York}, \bibinfo{address}{New York, NY}, \bibinfo{pages}{167--189}.
\newblock
\showISBNx{9781493903788}
\urldef\tempurl%
\url{https://doi.org/10.1007/978-1-4939-0378-8\_8}
\showDOI{\tempurl}


\end{thebibliography}

\appendix

\end{document}